\documentclass[sigconf, nonacm]{acmart}
\usepackage{subfiles}
\usepackage{xr}
\renewcommand\footnotetextcopyrightpermission[1]{}
\usepackage{mathtools}
\usepackage{enumitem}
\usepackage{mathtools}
\usepackage{tikz}
\usepackage{pifont}
\usepackage{multirow}
\usepackage{booktabs,arydshln}
\usepackage{colortbl}
\usepackage{bigdelim}
\usepackage{tcolorbox}
\usepackage{subcaption}
\usepackage{amsmath}
\usepackage{xparse}
\usepackage[normalem]{ulem}
\usepackage{soul}

\NewDocumentCommand{\dn}{e{_^}}{%
  _{\IfValueT{#1}{#1}\vphantom{\smash[b]{|}}}
  ^{\IfValueT{#2}{#2}\vphantom{\smash[t]{\big|}}}
}

\usepackage[ruled,linesnumbered, lined, boxed, commentsnumbered]{algorithm2e}
\usepackage{algorithmic}
\usepackage{multirow}
\usepackage{subcaption}
\usepackage{enumitem}

\newcommand\vldbdoi{XX.XX/XXX.XX}

\newcommand\vldbavailabilityurl{URL_TO_YOUR_ARTIFACTS}
 
\usepackage{color}
\definecolor{lightgray}{gray}{0.75}
\usepackage{tcolorbox}
\newtcolorbox{mybox}[3][]
{
  colframe = #2!25,
  colback  = #2!10,
  coltitle = #2!20!black,  
  title    = {#3},
  #1,
}

\usepackage{cleveref}

\newcounter{reviewer}

\makeatletter
\newcommand{\resplabel}[1]{\label[response]{resp:#1}}
\makeatother

\crefname{response}{Our Response}{Our Responses}

\newcounter{metarev}[meta]

\newcounter{revcomment}[reviewer]

\newcounter{reveval}[reviewer]

\newcounter{revrev}[reviewer]

\newcounter{response}[reviewer]

\renewcommand\theresponse{\thereviewer.\arabic{response}}

\makeatletter
\def\adl@drawiv#1#2#3{%
        \hskip.5\tabcolsep
        \xleaders#3{#2.5\@tempdimb #1{1}#2.5\@tempdimb}%
                #2\z@ plus1fil minus1fil\relax
        \hskip.5\tabcolsep}
\newcommand{\cdashlinelr}[1]{%
  \noalign{\vskip\aboverulesep
           \global\let\@dashdrawstore\adl@draw
           \global\let\adl@draw\adl@drawiv}
  \cdashline{#1}
  \noalign{\global\let\adl@draw\@dashdrawstore
           \vskip\belowrulesep}}
\makeatother

\AtBeginDocument{%
  \providecommand\BibTeX{{%
    \normalfont B\kern-0.5em{\scshape i\kern-0.25em b}\kern-0.8em\TeX}}}

\definecolor{dred}{HTML}{B85450}
\definecolor{dblue}{HTML}{6C8EBF}
\definecolor{dyellow}{HTML}{D79B00}
\definecolor{dpurple}{HTML}{9673A6}
\definecolor{dgreen}{HTML}{82B366}
\definecolor{tgreen}{HTML}{D5E8D4}
\definecolor{tpurple}{HTML}{E1D5E7}
\definecolor{fgray}{HTML}{666666}
\definecolor{bgray}{HTML}{F5F5F5}
\definecolor{bblue}{HTML}{EAF1FC}

\makeatletter
\newcommand{\subalign}[1]{%
  \vcenter{%
    \Let@ \restore@math@cr \default@tag
    \baselineskip\fontdimen10 \scriptfont\tw@
    \advance\baselineskip\fontdimen12 \scriptfont\tw@
    \lineskip\thr@@\fontdimen8 \scriptfont\thr@@
    \lineskiplimit\lineskip
    \ialign{\hfil$\m@th\scriptstyle##$&$\m@th\scriptstyle{}##$\hfil\crcr
      #1\crcr
    }%
  }%
}
\makeatother

\definecolor{Plum}{HTML}{C2938D}
\usepackage[colorinlistoftodos,prependcaption,textsize=tiny]{todonotes}
\usepackage{varwidth}
\usepackage{marginnote}

\newcommand{\RII}[1]{{}}

\definecolor{amethyst}{rgb}{0.6, 0.4, 0.8}

\newcommand{\delete}[1]{}
\newcommand{\techrep}[1]{}

\newcommand{\revision}[1]{{#1}}

\begin{document}

\newcommand{\name}{Biathlon\xspace}
\newcommand{\AGG}{{\theta}}
\newcommand{\Scaling}{\kappa}
\newcommand{\Ufeatures}{U_x}
\newcommand{\Uinference}{U_y}
\newcommand{\ConfLevel}{\tau}
\newcommand{\ErrorBound}{\delta}
\newcommand{\InitRatio}{\alpha}
\newcommand{\StepSize}{\gamma}
\newcommand{\MOp}{{\mathcal{M}}}
\newcommand{\Lat}{\mathcal{L}}

\newcommand{\NWorkload}{seven\xspace}

\newcommand{\MinSpeedup}{5.3$\times$\xspace}
\newcommand{\MaxSpeedup}{16.6$\times$\xspace}
\newcommand{\MinAccLoss}{0\%\xspace}
\newcommand{\MaxAccLoss}{1\%\xspace}

\newcommand{\Trips}{Trips-Fare\xspace}
\newcommand{\CheapTrips}{Trips-Cheap\xspace}
\newcommand{\TickPrice}{Tick-Price\xspace}
\newcommand{\TickPricePlus}{Tick-Price+\xspace}
\newcommand{\Bearing}{Bearing-Imbalance\xspace}
\newcommand{\BearingPlus}{Bearing-Imbalance+\xspace}

\newcommand{\ScoreCls}{F1-score\xspace}
\newcommand{\ScoreReg}{$r^2$-score\xspace}

\newcommand{\DefaultInitRatio}{0.05\xspace}
\newcommand{\DefaultStepRatio}{0.01\xspace}
\newcommand{\DefaultM}{1000\xspace}

\newcommand{\DefaultConfLevel}{0.95\xspace}
\newcommand{\DefaultErrorBoundTrips}{1.5\xspace}
\newcommand{\DefaultErrorBoundTick}{0.04\xspace}
\newcommand{\DefaultErrorBoundBattery}{186.7\xspace}
\newcommand{\DefaultErrorBoundTurbofan}{4.88\xspace}

\newcommand{\AvgIterTrips}{2.75\xspace}
\newcommand{\AvgIterTick}{1.26\xspace}
\newcommand{\AvgIterBattery}{3.87\xspace}
\newcommand{\AvgIterTurbofan}{1.41\xspace}
\newcommand{\AvgIterTDFraud}{4.27\xspace}
\newcommand{\AvgIterMachinery}{2.75\xspace}
\newcommand{\AvgIterStudent}{3.65\xspace}

\newcommand{\AvgSamplesTrips}{14.1\%\xspace}
\newcommand{\AvgSamplesTick}{5.6\%\xspace}
\newcommand{\AvgSamplesBattery}{7.9\%\xspace}
\newcommand{\AvgSamplesTurbofan}{5.4\%\xspace}
\newcommand{\AvgSamplesTDFraud}{11.1\%\xspace}
\newcommand{\AvgSamplesMachinery}{6.7\%\xspace}
\newcommand{\AvgSamplesStudent}{14.4\%\xspace}

\setcounter{figure}{0}
\setcounter{table}{0}

\title{Biathlon: Harnessing Model Resilience for Accelerating ML Inference Pipelines}

\author{Chaokun Chang  \quad  Eric Lo  \quad Chunxiao Ye}
\affiliation{%
  \institution{The Chinese University of Hong Kong}
}

\begin{abstract}
Machine learning inference pipelines commonly encountered in data science and industries often require real-time responsiveness due to their user-facing nature. However, meeting this requirement becomes particularly challenging when certain input features require aggregating a large volume of data online. Recent literature on interpretable machine learning reveals that most machine learning models exhibit a notable degree of resilience to variations in input. This suggests that machine learning models can effectively accommodate approximate input features with minimal discernible impact on accuracy.
In this paper, we introduce \name, a novel ML serving system that leverages the inherent resilience of models and determines the optimal degree of approximation for each aggregation feature. This approach enables maximum speedup while ensuring a guaranteed bound on accuracy loss. We evaluate \name on real pipelines from both industry applications and data science competitions, demonstrating its ability to meet real-time latency requirements by achieving \MinSpeedup to \MaxSpeedup speedup 
with almost no accuracy loss.
\end{abstract}
\maketitle


\begingroup
\renewcommand\thefootnote{}\footnote{\noindent
This work is licensed under the Creative Commons BY-NC-ND 4.0 International License. Visit \url{https://creativecommons.org/licenses/by-nc-nd/4.0/} to view a copy of this license. For any use beyond those covered by this license, obtain permission by emailing \href{mailto:info@vldb.org}{info@vldb.org}. Copyright is held by the owner/author(s). Publication rights licensed to the VLDB Endowment. \\
}\addtocounter{footnote}{-1}\endgroup

\ifdefempty{\vldbavailabilityurl}{}{
\vspace{.3cm}
\begingroup\small\noindent\raggedright\textbf{PVLDB Artifact Availability:}\\
The source code, data, and/or other artifacts have been made available at \url{https://github.com/ChaokunChang/Biathlon}.
\endgroup
} 

\section{Introduction} 

Machine Learning (ML) has gained traction across a diverse array of applications. In the training phase, developers gather data to train their machine learning models.
In the serving phase, 
the trained model is deployed within an \emph{inference pipeline},
which accepts user inputs and carries out real-time model inference.

A typical real-time inference pipeline consists of a series of operations related to \emph{feature preparation}. Generally, there are multiple feature preparation operators responsible for generating features based on runtime inputs. Once all the features are prepared, the model inference operator processes these features as input and produces an inference result that is returned to the user.
Despite recent advancements in deep learning,
traditional models like Linear Regression, Decision Tree, XGBoost still demonstrate exceptional performance and accuracy
on tabular and data science data \cite{JoinBoost}.
In fact, most inference pipelines in Kaggle \cite{kaggle}
are using traditional models like random forests and gradient boosting \cite{JoinBoost,WhyTreeModel,DSFound,Chen_Li_IML}.
Traditional models are lightweight in terms of inference cost \cite{Willump}.
The heavy-lifting part of those pipelines, however, often falls on the feature preparation operators 
when they need to aggregate a large volume of data
\cite{OpenMLDB,FEBench}.

Recent literature in the field of machine learning interpretation \cite{IMLBook,LIME,Anchor} shows that machine learning models exhibit a notable degree of \textbf{resilience} to variations in input. This phenomenon implies that the inference results produced by machine learning models often demonstrate a certain level of stability, even in the presence of imprecise input features. This suggests that machine learning models are capable of accommodating {\bf approximate input features} with {\bf minimal discernible impact on their predictive accuracy}.

Approximately computing input features can yield significant acceleration to an entire inference pipeline by alleviating the data processing burden at its core. Sampling-based approximation query processing (AQP) enables the preparation of aggregation features using a smaller subset of the original dataset \cite{AQPGuoliang, BlinkDB, VerdictDB, SampleSeek}. Consequently, this approach offers a powerful means of expediting the feature preparation process within an inference pipeline.
However, this endeavor is particularly challenging. 
First, industrial and data science inference pipelines
    typically involve multiple input features, each exerting a non-uniform impact on the final inference result. Consequently, determining the appropriate approximation level for each feature becomes a non-trivial task, as their respective influences vary.
Second, inference pipelines often comprise a complex flow of inter-dependent operators. Navigating this intricate interplay engenders further complexity when deciding the appropriate approximation level for each feature. The ultimate goal is to strike a delicate balance between maximizing speedup through approximation while maintaining an acceptable level of accuracy loss within permissible bounds.

We propose \name, a new ML real-time serving system that harnesses the resilience of ML models to accelerate the execution of inference pipelines. 
\name effectively determines the appropriate approximation degree for each feature by considering both its computational cost and its importance in relation to the current inference result. When a feature imposes a high processing burden, \name allocates a higher level of approximation to expedite execution. Conversely, if a feature significantly impacts the inference result, indicating sensitivity to variations in that feature, \name prescribes a lower approximation level to preserve accuracy.

\begin{figure}
  \centering
  \includegraphics[width=0.5\linewidth]{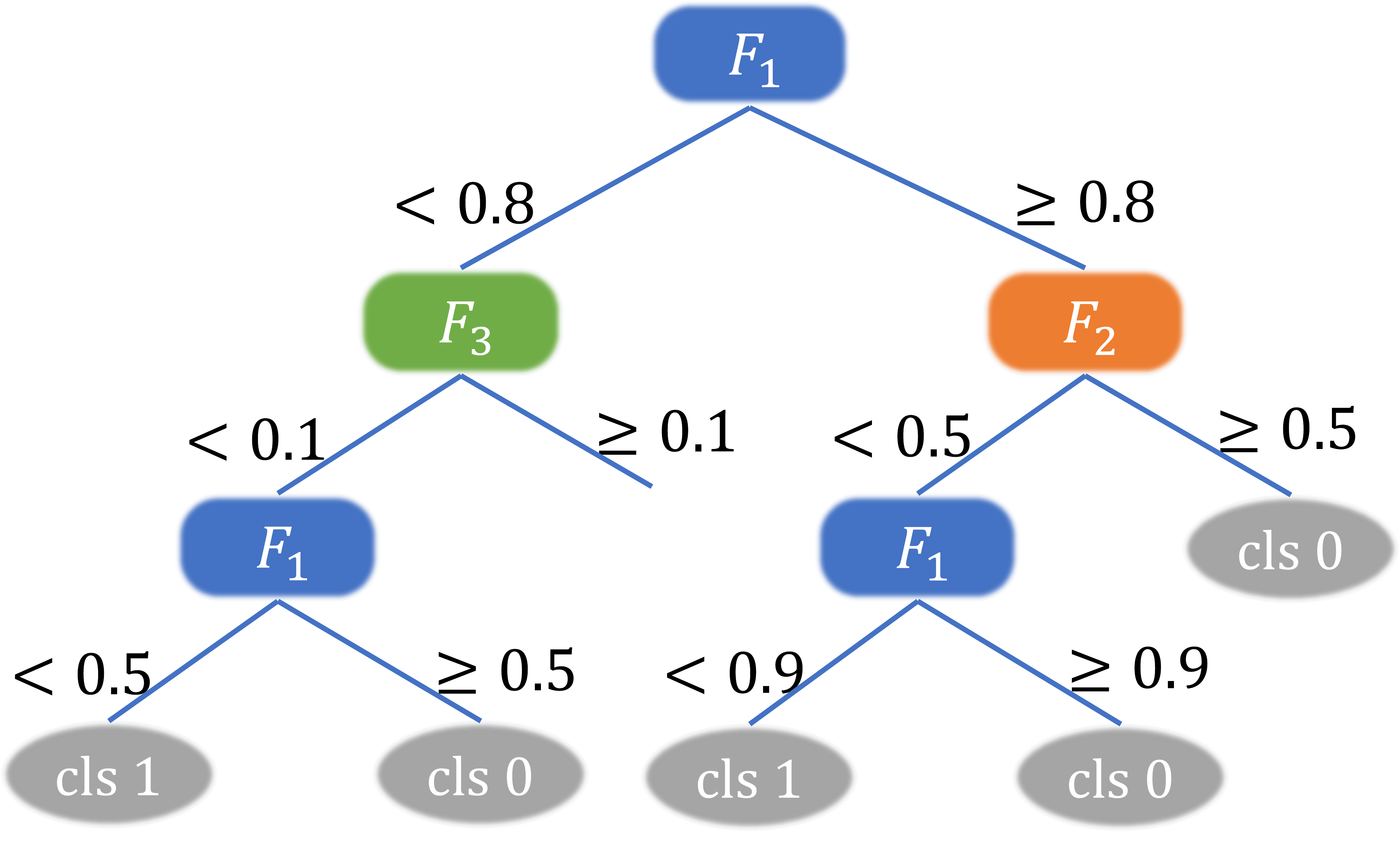}
  \caption{Decision tree example}
  \vspace{-1.5em}
  \label{fig:dtmodel}
\end{figure}

It is imperative to note the importance of a feature is 
\emph{input-sensitive}.
Specifically, the importance of a feature fluctuates with different input values. This variability stems from the dynamic interaction between features across varied input values. For instance, 
consider the decision tree in Figure \ref{fig:dtmodel}.  The importance of feature $F_3$ 
depends on the value of feature $F_1$.  When $F_1 \ge 0.8$, $F_3$ becomes immaterial.
Conversely, when $F_1 < 0.8$, $F_3$ becomes very important.
As the importance of a feature 
to a specific inference result 
hinges upon its runtime input value, it is implausible to ascertain an approximation degree for each feature offline.
Hence, \name uses \RII{Responses 2.2} an \revision{\emph{online}} approach to determine the \emph{approximation plan} during runtime for each individual inference pipeline execution.
Specifically, \name incrementally extracts samples to approximate the feature values and hence the model output.
During the process, it also estimates the importance of features using the samples. By utilizing these estimates, \name incrementally refines the approximation plan to draw more samples and early-stops once the model output is statistically guaranteed to be correct.

\name exhibits broad applicability across various ML models. For models with discrete output (e.g., classification), \name provides a probabilistic guarantee that the inference result obtained using its approach is identical to the inference result derived using exact features. For models with continuous output (e.g., regression), \name ensures a probabilistic guarantee that the inference result lies within a bounded error relative to the inference outcome produced with exact features. 
This characteristic makes \name a versatile solution for enhancing the performance of a wide variety of ML inference pipelines.
To demonstrate the efficacy of \name, we conducted a comprehensive evaluation on real inference pipelines, originating from both industry applications and data science competitions. \name successfully harnesses the pipeline model
resiliency to offer \MinSpeedup and \MaxSpeedup speedup, without noticeable degradation in accuracy.

\section{Background}
A machine learning inference pipeline is often a workflow
of  operators collective 
for feature preparation
and model inference.
Typical operators include:

\begin{enumerate}[leftmargin=1.5em]
    \item \textit{Datastore Operators}: 
    These operators involve external data access such as querying a database. 
    Some datastore operators are lightweight, particularly when suitable indexes are available (e.g., retrieving the gender of a user based on their unique user ID). 
    Some are heavyweight, 
    requiring the retrieval of significant amounts of data for aggregation
    (e.g., 
    counting the number of clicks in
    a user group that shares common interests with the current user).    
    
    \item \textit{Transformation Operators}: 
    These operators are responsible for transforming the data.
      They require no external 
    data access and are lightweight.  Examples of such transformations include 
    Standard Scaling, One-Hot Encoding, and N-gram.
    
    \item \textit{Model Inference Operators}: 
    A model inference operator is the terminal operator 
    that ends a pipeline and generates a prediction result.
    Inferences based on traditional models such as Linear Regression (LR), Support Vector Machines (SVM), and tree-based models like XGBoost and LightGBM are not computationally expensive \cite{Willump}. Our experiments demonstrate that model inference operations typically execute in milliseconds, falling within the same ballpark as lightweight datastore lookup and data transformations.     
    
\end{enumerate}

\begin{figure}
  \centering
  \includegraphics[width=\linewidth]{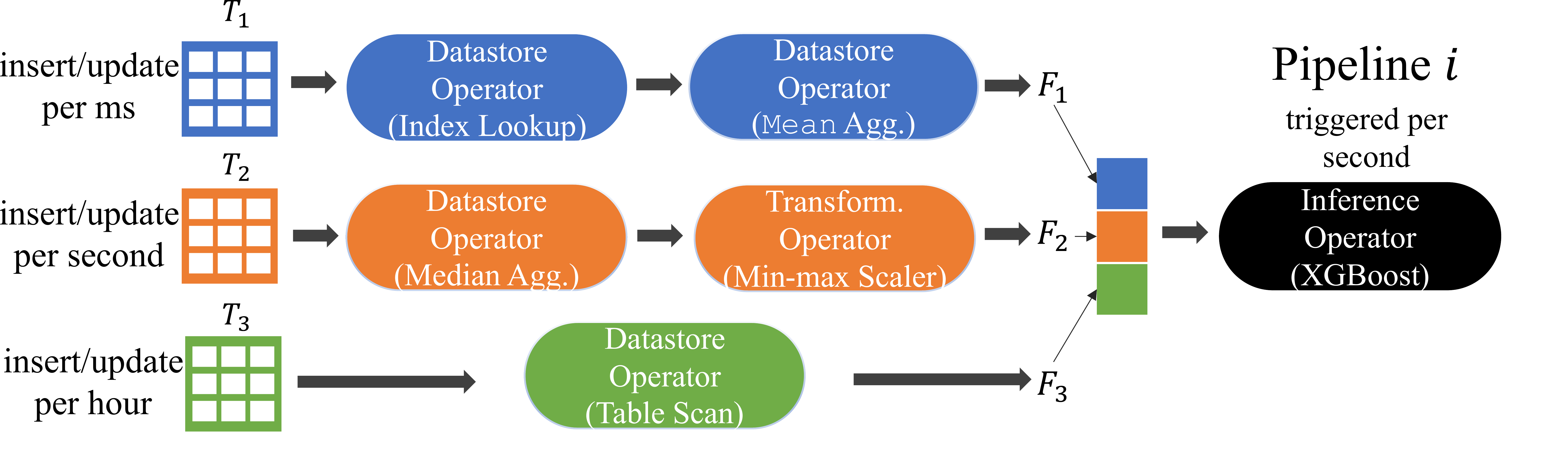}
  \caption{A (simplified) inference pipeline from Kaggle}
  \label{fig:ipexample}
\end{figure}

\autoref{fig:ipexample} shows an inference pipeline from Kaggle \cite{kaggle}, simplified for illustration purpose. 
This example pipeline consists of five feature preparation operators collectively forming three feature preparation sub-pipelines, each yielding a distinct feature utilized in model inference.
The data tables in the pipeline have different update frequencies.

To address the aggregation bottleneck in real-time inference pipelines, industries such as Databricks\cite{Databricks-Feature-Store}, Vertex AI \cite{Vertext-AI-Feature-Store}, and Tecton \cite{Feathr} often pre-aggregates some features offline. These pre-aggregated features are stored in specialized databases commonly known as ``feature stores'' 
\cite{Feathr,Feast,Feathub,Hopsworks}
for subsequent online usage. \RII{Responses 1.2, 1.3} However, the utilization of feature stores inherently introduces 
\revision{space overhead and
a certain degree of staleness to the features.
This staleness can potentially result in unbounded errors in the inference results.

RALF \cite{RALF} is an optimized feature store.
To reduce errors resulting from potentially stale features, it periodically selects a subset of features to refresh based on a cost budget.
RALF assumes the error of each prediction can be promptly obtained and leverages those errors to establish a feedback loop, determining when to reuse a cached feature and when to refresh and recompute a feature. 
Unfortunately, not many ML pipelines can obtain the error of each prediction promptly.
For instance, in the Trip-Fare pipeline we used in our experiments, 
the error of a trip-fare prediction can only be obtained after the trip has concluded, which may take minutes or even hours. In such cases, RALF often fails to establish an effective feedback loop due to lagged information,
resulting in noticeable accuracy loss caused by stale features.
}

Willump \cite{Willump} exploits the statistical properties of machine learning models within inference pipelines to reduce the cost of feature preparation. Willump constructs and utilizes an approximate model for simple inputs,
and only uses the original model for complex inputs.
The approximate model requires fewer features as inputs, 
directly cutting the cost of feature preparation.
However, training the approximation model may pose challenges as it requires access to the training set, which voids all use cases
whose training data are not available (e.g., the use of 
pre-trained models).

\name distinguishes itself as a pioneer in accelerating inference pipelines by taking a different approach from Willump. Instead of approximating the model, \name focuses on approximating the features. 
Approximating the computation of expensive aggregates 
belongs to a well-established topic called Approximate Query Processing (AQP).
The key, however, lies in determining the appropriate level of approximation for each feature, balancing between maximum speedup and minimal prediction accuracy loss.
\name leverages the inherent resilience found in machine learning models to address this challenge.

\subsection{Approximate Query Processing}

Approximate Query Processing (AQP) is a prominent technique employed to swiftly return approximate responses for queries necessitating the processing of a large volume of data \cite{AQPGuoliang, BlinkDB, VerdictDB, SampleSeek}. 

Sampling has been the most prevalent AQP approach \cite{BlinkDB,VerdictDB,XDB,SampleSeek,Quickr,DBO,OLA}, owing to its three pivotal characteristics: (1) \emph{generality}, enabling its broad applicability across a diverse spectrum of aggregation operators encompassing distributive and holistic aggregation; (2) \emph{simple}, requiring solely the specification of a sample size to work; and (3) \emph{theoretical guarantees}, providing bounds to the approximation results.

Sampling-based AQP methods are predominantly dichotomized based on the employed sampling algorithm.
    AQP based on \emph{uniform sampling} \cite{OLA,DBO,GOLA,SampleSeek,Quickr,XDB} 
    selects samples in a randomized fashion. This method offers several advantages: it is workload-independent and necessitates no data preprocessing.
    On the contrary, AQP based on \textit{biased sampling} \cite{BlinkDB,VerdictDB} 
    selects samples based on historical workloads, endowing certain records with a higher likelihood of selection. While 
    biased sampling typically demands a smaller number of samples,
    it is constrained by its reliance on past workloads and susceptibility to workload shifts.

\textit{Online Aggregation} \cite{DBO,OLA} incrementally draw samples until the estimated error of the aggregation result attains the specified accuracy target.  
\revision{Online aggregation can support
standard statistics like 
{\tt SUM},
{\tt COUNT}, 
{\tt AVG},
{\tt VAR}, {\tt STD}, {\tt MEDIAN}, and {\tt QUANTILE}.
However, it cannot support \RII{Responses 0.2, 1.3}
{\tt TOP-K},
{\tt DISTINCT}, 
and extreme statistics 
{\tt MIN} and {\tt MAX}.}
Conversely, systems that draw (biased) samples offline based on historical workloads 
do not have an online sampling overhead
\cite{BlinkDB} and support more operators.
However, their approximated answers do not necessarily meet the user-specified accuracy target.

Recent advancements in AQP
train ML models to replace biased samples, 
resulting in improved estimation accuracy \cite{DBL,DeepDB,DBEst}. 
However, this approach is still prone to workload shifts
and agnostic to user-specified accuracy targets.
\name is an ML serving system that utilizes online aggregation 
to approximate expensive features.
Instead of specifying an accuracy target for the aggregation operators, \name specifies the accuracy target for the final prediction result and ``back-propagates'' this target to determine the accuracy targets for individual upstream aggregate operators
by considering their importance and processing costs.

\subsection{Feature Importance}

In machine learning, Feature Importance plays a significant role in various aspects such as Feature Selection, ML Interpretability, and ML security. For instance, the concept of ``permutation importance'' \cite{sklearn} is commonly used to measure the importance of a feature. It is defined as the decrease in a model's score when the values of that feature are randomly shuffled. By permuting the feature values, the relationship between the feature and the target is disrupted, and the resulting drop in the model score indicates the extent to which the model relies on that particular feature.

In Feature Selection, models can be built using only features with positive importance scores \cite{PermutationImp}. In Explainable AI, feature importance scores can be used to interpret specific inference outcomes derived from machine learning models \cite{Shapley, SHAP, LIME, Anchor}. In Adversarial Machine Learning, evasion attacks \cite{FI4Attack} involve carefully preparing adversarial examples to cause mis-classification during inference time. Feature importance can guide the search for adversarial examples by identifying  critical features \cite{FI4AdvExample}.

To the best of our knowledge, \name represents a pioneering effort in utilizing feature importance to derive approximation plans for accelerating the execution of inference pipelines.

\section{\name}
Given an unoptimized inference pipeline $G$, pipeline inputs (e.g., user ID), an error bound $\ErrorBound$, and a confidence level $\ConfLevel$, the goal of \name is to execute $G$ to obtain an inference result $\hat{y}$ that satisfies the accuracy guarantee specified in \autoref{eq:accuracy_target}, with minimal execution cost:
    \begin{equation} \label{eq:accuracy_target}
        Pr(|Y - \hat{y}| \le \ErrorBound) \ge \ConfLevel
    \end{equation}
where $Y$ represents the inference result without \name's optimization (i.e. computing all features exactly). 

It is noteworthy that for classification pipelines, the value $\ErrorBound$ must be 0.
\autoref{eq:accuracy_target} intuitively guarantees that the deviation of the inference result obtained using \name from the actual inference result will not exceed $\ErrorBound$ with at least $\ConfLevel$ confidence.
The execution cost of the inference pipeline is:
\vspace{-1em}
\begin{equation}
    C^{z} = \|z\|_1 = \sum_{1}^{k} z_j
\end{equation}
where $z=[z_1, \cdots, z_k]$ is the \emph{approximation plan},
a vector that denotes the sample size  $z_j$ for each feature, with $z_j$ not 
exceeding the total number of records $N_j$ for that feature, i.e. $0 \le z_j \le N_j$. 

Currently, \name only approximates features that are computed by expensive aggregation operators. We do not approximate other operators (e.g., scaling) 
because their cost is relatively low compared to aggregation.
\revision{\RII{Responses 0.2, 1.3}
Inheriting the limitation from online aggregation \cite{OLA,DBO,GOLA,SampleSeek,Quickr,XDB,KnowingWrong}, \name does not approximate  
{\tt TOP-K},
{\tt DISTINCT}, 
{\tt MIN}, and {\tt MAX}.}
So, $k$ is the number of aggregation features approximated by \name,
and $C^z$ is the cost of executing the pipeline according to $z$,
measured in terms of the total number of samples across all aggregation features.

The optimal approximation plan $z^*$ in \name is the one that satisfies \autoref{eq:accuracy_target} with the minimal cost $C^*$. However, similar to many query optimization problems, finding the optimal plan outright without knowing the exact inference result $Y$ is infeasible. Consequently, \name adopts an iterative algorithm to progressively approach \autoref{eq:accuracy_target} step by step. 

\subsection{Workflow of \name}
\begin{figure}
    \centering
    \includegraphics[width=0.85\linewidth]{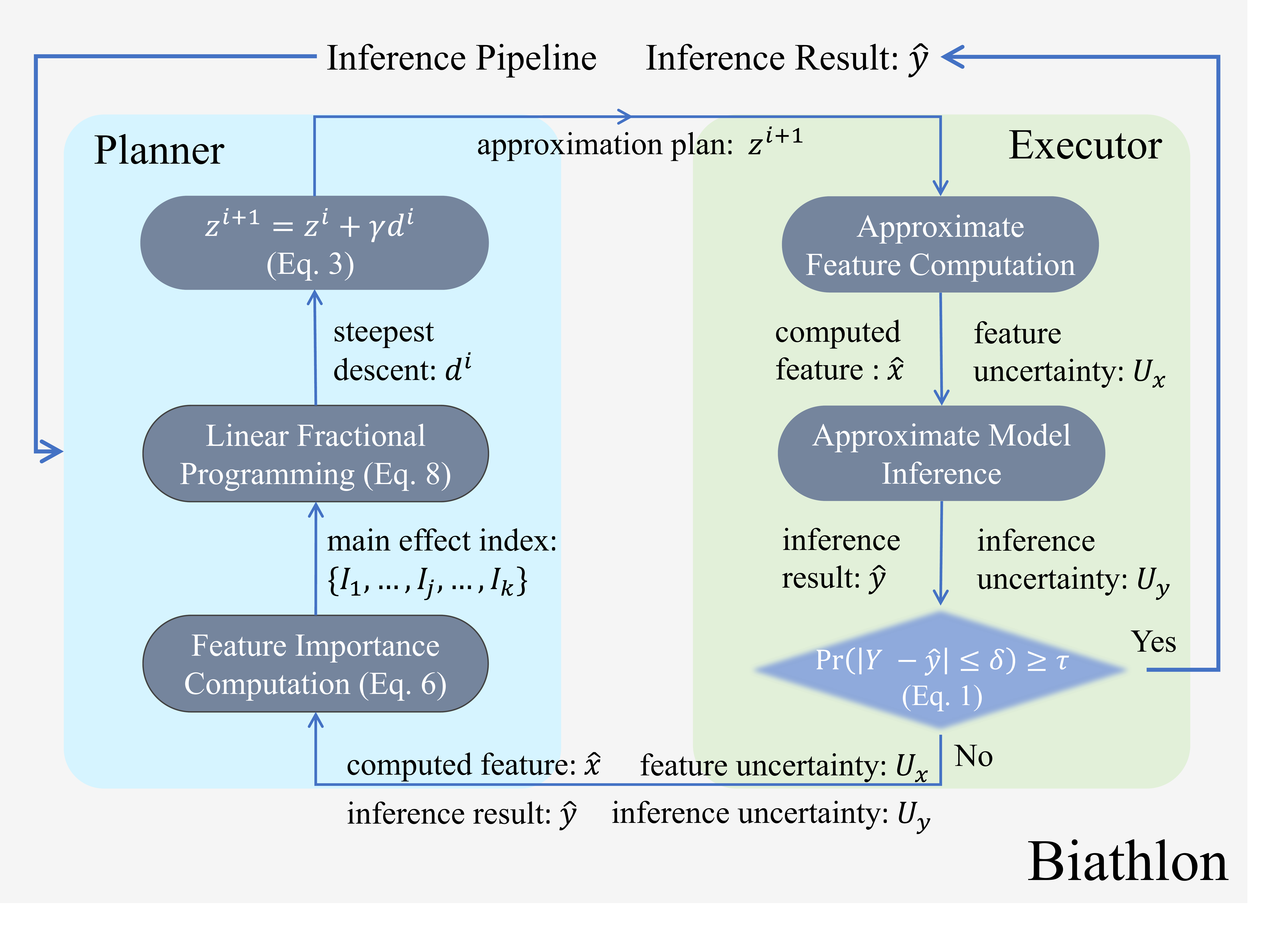}
    \vspace{-1em}
    \caption{System Overview of \name}
    \vspace{-1em}
    \label{fig:overview}
\end{figure}

\autoref{fig:overview} illustrates the workflow of \name.  
\name comprises two components: (1) The \emph{Planner}, responsible for devising an approximation plan \revision{online} \RII{Response 2.2} for the execution of the inference pipeline, and (2) The \emph{Executor}, tasked with executing the inference pipeline approximately in accordance with the plan proposed by the Planner.

\name, when receiving an inference request from a user, begins with the Planner formulating an initial plan $z^0$ of initial samples for each input feature. The Executor then utilizes this plan for execution.

The execution process within the Executor can be divided into two stages: Approximate Feature Computation (AFC) and Approximate Model Inference (AMI). During the AFC stage, the Executor computes the values of approximate features $\hat{x}$ and estimates their \textbf{feature uncertainties} $\Ufeatures$. In the subsequent AMI stage, the Executor performs model inference using the approximate features $\hat{x}$ to obtain the approximate inference result $\hat{y}$ and estimates its \textbf{inference uncertainty} $\Uinference$.
Subsequently, \name performs a validation check to determine whether the current inference result $\hat{y}$ aligns with the specified requirement in \autoref{eq:accuracy_target}. 
\revision{
Specifically, given the inference uncertainty $U_y$, we can calculate the cumulative probability \RII{Response 2.2} that $U_y$ falls within the error interval $(-\delta, \delta)$.  If the cumulative probability area within $(-\delta, \delta)$ is greater than or equal to $\ConfLevel$, then}
the condition in \autoref{eq:accuracy_target} is met, and \name would return the approximate inference result $\hat{y}$ to the users. Otherwise, \name initiates a feedback loop and channels ($\hat{x}$, $\Ufeatures$) and ($\hat{y}$, $\Uinference$) back to the Planner to devise a new approximation plan $z^1$ for the next iteration of execution. \name continues iterating through the feedback loop \revision{and draw \RII{Response 2.3} more samples}
until the user obtains an inference result whose inference uncertainty meets  \autoref{eq:accuracy_target}.
\revision{
In other words, although unlikely, \name may need to draw all samples to compute the exact feature \RII{Response 2.3} in order to satisfy Equation \ref{eq:accuracy_target} when confronted with worst-case scenarios
(e.g., malicious data distributions).
}

\subsection{Approximate Feature Computation (AFC)}

In this stage, \name calculates the values of the features. For non-targeting features, \name computes their exact values. In the case of targeting aggregation feature $j$, \name operates similarly to online aggregation, providing efficient estimations for the aggregation values through a three-step process.

First, \name randomly selects a sample $S_j$ 
of size $z_j$ for attribute $j$ according to the approximation plan $z$. 
Next, \name estimates the approximate value of feature $j$ using sample $S_j$. This process resembles existing sampling-based AQP techniques. Initially, the aggregation operator $\AGG_j$ is applied to its input from the selected sample $S_j$. The resulting aggregation is then scaled to obtain an estimate of the true aggregation value on the entire dataset, denoted as $\hat{x}_j = \Scaling_j (\AGG_j(S_j))$. Here, $\Scaling_j$ represents the scaling operator specific to feature $j$.  Lastly, \name estimates the uncertainties $\Ufeatures$ of the approximated features.

In contrast to traditional AQP systems that employ statistics (e.g., standard deviation) to quantify result uncertainty, \name directly employs the error distribution between the approximate feature and the exact feature to capture the estimation uncertainty $\Ufeatures$. This approach is chosen because, unlike online aggregation, the approximation results here serve as intermediate results rather than final results. Therefore, it aims to preserve as much information as possible for estimating the inference uncertainty $\Uinference$ later.

In \name, 
\revision{for standard conditional aggregation operators that can be supported by AQP, \RII{Response 2.1}
including \texttt{SUM}, \texttt{COUNT}, \texttt{AVG}, \texttt{VAR}, and \texttt{STD}}, 
we adhere to the approach proposed in \cite{AQPHandbook} by setting the error distribution of the approximate aggregation $\Ufeatures$ as a normal distribution.
Consequently, estimating $\Ufeatures$ involves finding the mean $\mu$ and standard deviation $\sigma$ of the error distribution. 
The mean $\mu$ is 0 since sampling-based AQP can provide unbiased estimation. 
\revision{For holistic measures like \texttt{MEDIAN} and \texttt{QUANTILE}, \RII{Response 2.1} we use Empirical Bootstrap \cite{Bootstrap, RDBQuerywithBootstrap} to obtain an empirical distribution}.

Online aggregation eliminates the need for data pre-processing, enabling \name to handle very dynamic data. Furthermore, online aggregation draws samples incrementally, which avoids repeated data access when AFC is triggered multiple times with different approximation plans before \name stops.
With this design, if \name is unsatisfied 
with the inference result and the planner suggests to
increase the sample size of a selected feature from 
from $z_j$ to $z_j'$, the Executor can incrementally draw ($z_j' - z_j$) new samples instead of drawing $z_j'$ samples from scratch.
This incremental computation mechanism effectively avoids redundant data access
and works for most aggregation operators, including both distributive measures and holistic measures \cite{HyperLogLog,IncWinAGG}.

\subsection{Approximate Model Inference (AMI)} \label{sec:AMI}
The AMI stage in \name serves a dual purpose: (1) computing the (approximate) inference result $\hat{y}$ using the approximate features 
and (2) estimating the uncertainty of the approximate inference result $\Uinference$. 

Computing the approximate inference result $\hat{y}$ is straightforward ---
\name directly incorporates the approximate feature values $\hat{x}$ into the model inference operator to derive the approximate inference result: 
$\hat{y} = \MOp(\hat{x})$, where $\MOp$ represents the model inference operator.
In \name, the error of the inference result refers to the discrepancy between the approximate and the exact inference results.
Hence, given the uncertainty of input features $\Ufeatures$,
estimating the uncertainty of inference result $\Uinference$
is actually a problem known as \emph{uncertainty propagation} (UP) \cite{UPProblem}.

There are two types of methods to solve the UP problem: analytical methods and black-box methods. The former is contingent upon the availability of model-specific closed-form formulas, limiting its applicability to simple models like Linear Regression and rendering it unsuitable for \name's objective of supporting a diverse array of models. Consequently, \name addresses the UP problem through a black-box method based on Monte Carlo simulations (MCS).
While alternative black-box methods \cite{BBUP} exist, they lack the flexibility of MCS and often suffer from the curse of dimensionality. 
Standard Monte Carlo methods, however, are computationally intensive due to sampling inefficiency. Hence, \name employs \textit{quasi-Monte Carlo (QMC)} \cite{MCQMC}, harnessing low-discrepancy sequences to uniformly cover the input space, thereby achieving comparable estimation accuracy with fewer samples.

Based on QMC, \name estimates the uncertainty of the inference result $\Uinference$ in four steps.

\begin{enumerate}[leftmargin=1.5em] 
\item  Generate $m$ i.i.d. feature samples $x^1, \cdots, x^m$, 
with 
$x^i = \Ufeatures + \hat{x}$.
Each $x^i$ still follows a normal distribution as $\Ufeatures$.
    Note that the $m$ feature samples are generated using a low-discrepancy sequence \cite{SobolSequence}, also referred to as a quasi-random sequence, to achieve fast convergence. 

\item Conduct model inference on the generated feature samples, yielding $m$ inference samples $y^1, \cdots, y^m$, where $y^i = \MOp(x^i)$.

\item Model the distribution of the true inference result $Y$ based on the ensemble of $m$ inference samples. 
    In the case of a regression model, 
    the distribution of $Y$ follows a normal distribution $N(\bar{y}, \sigma_y^2)$, where $\bar{y} = E(Y) \simeq \frac{1}{m} \sum_{i=1}^{m} y^i$, and $\sigma_y^2 = E((Y - \bar{y})^2) \simeq \frac{1}{m} \sum_{i=1}^{m} (y^i - \hat{y})^2 $. 
    Alternatively, if the model is a classification model, the distribution of $Y$ is a categorical distribution, requiring estimation of the probabilities of all possible classes using their frequencies in the inference samples. The probability of class $j$ is estimated as $p_j = \frac{1}{m} \sum_{i=1}^{m} \mathcal{I}_{y^i = j}$, where $\mathcal{I}$ is an indicator function that $\mathcal{I}_{y = j} = 1$ when $y=j$ and  $\mathcal{I}_{y = j} = 0$ otherwise.

\item Compute the uncertainty, i.e. $\Uinference$.
    By definition, $\Uinference = Y - \hat{y}$.
    Hence, for regression models, $\Uinference$ follows a normal distribution $\Uinference \sim N(\bar{y} - \hat{y}, \sigma_y^2)$. On the other hand, for classification models, $\Uinference$ follows a Bernoulli distribution $\Uinference \sim Bernoulli\left(1 - p_{\hat{y}} \right)$, where $p_{\hat{y}}$ denotes the probability of class $\hat{y}$, i.e., $p_{\hat{y}} = \frac{1}{m} \sum_{i=1}^{m} \mathcal{I}_{y^i = \hat{y}}$.
\end{enumerate}

It is worth highlighting that Monte Carlo
methods exhibit a high degree of parallelizability in their computation.
\name leverages this property by performing the $m$ model inferences simultaneously in parallel. This approach effectively enables efficient estimation of $\Uinference$ 
with reduced time requirements.
As a side note, \name typically employs parametric methods to model the probability distribution of $Y$.  However, if $Y$ deviates from the distribution assumption of parametric methods, \name resorts to the use of non-parametric Kernel Density Estimation (KDE) instead.

\subsection{Planner}

The primary responsibility of the Planner in \name is to determine the approximation plan, denoted as $z$, at the beginning of each iteration. 
In the beginning,  
the Planner initializes the initial plan $z^0$ using a small percentage $\InitRatio$ of data records within each feature. Therefore, the initial plan is $z^0 = [\InitRatio N_1, \cdots, \InitRatio N_k]$, where $N_j$ represents the number of records for feature $j$.
For subsequent iterations $i > 0$, the Planner determines the next plan $z^{i+1}$ in accordance with \autoref{eq:policy}:
\begin{equation} \label{eq:policy}
    z^{i+1} = z^{i} + \gamma d^{i}
\end{equation}
\revision{where $d^{i}$ is a vector denoting the direction of the maximum
\RII{Response 2.3} 
reduction in inference uncertainty} based on the current plan $z^i$, and $\gamma$ denotes the \emph{step size}, governing the number of 
additional samples to allocate in each iteration.
Similar to any iterative optimization algorithm, the step size is a hyperparameter. An excessively small step size necessitates additional iterations to fulfill \autoref{eq:accuracy_target}, leading to increased overhead from more iterations. Conversely, an excessively large step size may result in an overshoot in terms of execution cost, causing \name to fulfill \autoref{eq:accuracy_target} using an excessively large number of unnecessary samples.

\newcommand{\Curd}{d^i}
\newcommand{\Dz}{\Delta{z}}
\newcommand{\Curz}{z^i}
\newcommand{\Nxtz}{\Curz + \Dz}
\newcommand{\CurVar}{Var(Y | \Curz)}
\newcommand{\NxtVar}{Var(Y | \Curz + \Dz)}
\newcommand{\VarReduction}{\CurVar - \NxtVar}

\revision{
The direction characterized by the 
maximum reduction in inference uncertainty $\Curd$ at $\Curz$  \RII{Response 2.3}
is as follows:
}

\begin{equation} \label{eq:sd-direction}
    \begin{aligned}
        \Curd = & \arg \max_{\Dz} \frac{\VarReduction}{\|\Dz\|_1}
    \end{aligned}
\end{equation}

\noindent
where $Var(Y | \Curz)$ represents the variance of the inference result when the approximation plan is $\Curz$, serving as a measure of the current level of inference uncertainty, which can be easily obtained given the inference uncertainty 
$U_y$  in AMI (Section \ref{sec:AMI}).

The vector $\Dz = [\Dz_1, \cdots, \Dz_k]$ specifies a direction for adjusting the current plan, with each $\Dz_j \in \{0, 1\}$ indicating how the sample size for feature $j$ should be modified. A value of $\Dz_j = 0$ signifies no change, while $\Dz_j = 1$ indicates the acquisition of $\gamma$ samples, considering the multiplication by the step size. It is pertinent to note that decreasing the sample size for a particular feature is not considered, 
as \name computes features incrementally, and the execution costs associated with a smaller sample size have already been accounted for in previous iterations. In addition, $\|\Dz\|_1$ is defined as $\|\Dz\|_1 = \sum_{j=1}^{k} \Dz_j$, reflecting the increase of execution cost in that direction.
Finally, we note that $\Curd$ is not differentiable as $z_i$ is discrete.

Directly computing $\Curd$ is not recommended due to the value of $\NxtVar$ depends on the execution result of the inference pipeline, requiring $2^k$ pipeline executions that 
include expensive aggregations and model inference.
Fortunately, we have been able to identify a shortcut to estimate $d^i$ with a closed-form solution.
Specifically, given a fixed increased sample $\Dz$, the corresponding variance reduction is related to the importance of the feature whose sample size is increased. 
The more important the feature is, the more the variance is reduced
by having more samples.
In machine learning, 
there are many measures to quantify the importance of a feature, including
LIME \cite{LIME}, Shapley Value \cite{Shapley}, SHAPE \cite{SHAP},
and Sobol Indices \cite{SobolIndices}.
Among those, we use Sobol Indices 
because they define feature importance exactly 
based on variance reduction.

Sobol Indices comprise a total of $2^k - 1$ indices with different orders ranging from first-order to $k$-th-order, where $k$ denotes the number of features. Among these, there are $k$ first-order indices $\{I_1, \cdots, I_j, \cdots, I_k\}$, also known as the \emph{Main Effect indices}, with $I_j$ measuring the importance of feature $j$. 
There are $\frac{k (k-1)}{2}$ second-order indices $\{ I_{12}, \cdots, I_{ij}, \cdots\}$, where $I_{ij}$ quantifies the importance of the interaction between features $i$ and $j$, and so forth for higher-order indices. In \name, 
the first-order Main Effect Indices are sufficient.
The Main Effect Index for feature $j$ is defined as 
\cite{SobolIndices}:

$$    I_j = \frac{Var_{X_j}(E_{\neg X_j}(Y|X_j))}{Var(Y)}$$

\newcommand{\JExactPlan}{z^i_{j*}}
\newcommand{\CurInd}{I^i}

where $X_j$ represents feature $j$, and $\neg X_j$ denotes all other features except $j$. 
The denominator represents the variance of the inference result, while the numerator represents the variance of the conditional expectation of $Y$ given $X_j$, quantifying the proportion of variance contributed by feature $j$.  
By the law of total variance, the numerator can also be seen as:
\vspace{-0.5em}
\begin{equation}\label{eq:lotv}
Var_{X_j}(E_{\neg X_j}(Y|X_j)) = Var(Y) - E_{X_j}(Var_{\neg X_j}(Y | X_j))    
\end{equation}

In our context, 
 the denominator of the main effect index for feature $j$, 
 i.e., the variance of the inference, is $\CurVar$.
The term $E_{X_j}(Var_{\neg X_j}(Y | X_j))$ 
in \autoref{eq:lotv}
represents the expectation of conditional inference variance given $X_j$, 
i.e., when $X_j$ is based on all $N_j$ records.
In our context, the denominator would then be:
\vspace{-0.5em}
$$\CurVar - E(Var(Y | \JExactPlan))$$

where $\JExactPlan$ is the plan $[z^i_1, \cdots, N_j, \cdots, z^i_k]$
with feature $j$ computed using all $N_j$ records.
Hence, putting it all together, the importance of feature $j$ 
at plan $z^i$ is:

\begin{equation}
      I_j^i = \frac{\CurVar - E(Var(Y | \JExactPlan))}{\CurVar}
\end{equation}

$I^i_j$ can also be computed efficiently
using the Sobol-Satelli method \cite{Saltelli-Method},
which is also QMC-based
like the one in AMI 
(Section \ref{sec:AMI}).
With those feature samples and inference results, 
we can derive $I^i_j$ for all $j$.

Utilizing the Sobol's Main Effect Index, we can estimate the variance reduction by summing the expected variance reduction caused by each feature. 
Let $\CurInd= [\CurInd_1, \cdots, \CurInd_k]$ denote the Sobol's Main Effect Index vector of all features based on the plan $z^i$.  Each individual 
$\CurInd_j$ indicates the expected contribution of feature $j$ to the inference variance reduction ratio if $j$ becomes exact.  Hence, the expected variance reduction would be $\CurVar \cdot  \CurInd_j$. 
Therefore, we can compute the unit reduction per future sample as $\frac{ \CurVar \cdot \CurInd_j}{N_j - z_j}$, and the variance reduction 
caused by giving the next iteration of samples to feature $j$ as $\frac{\CurVar \cdot \CurInd_j \cdot \Dz_j}{N_j - z_j}$. Hence, given  $\Dz$, we can estimate the overall inference variance reduction as $\sum_{j=1}^k \frac{\CurInd_j \Dz_j}{N_j - z_j} \CurVar$, i.e.
\vspace{-0.5em}
\begin{equation} \label{eq:var-red}
    \VarReduction \simeq (\frac{\CurInd}{N - z})^T \Dz \CurVar
\end{equation}

Consequently, we can transform Equation \ref{eq:sd-direction} into Equation \ref{eq:sd-cal}:

\begin{equation} \label{eq:sd-cal}
\begin{aligned}
    \Curd & \simeq \arg \max_{\Dz} {(\frac{\CurInd}{N - z})}^T \frac{\Dz}{\|\Dz\|_1} \CurVar \\
            & = \arg \max_{\Dz} {(\frac{\CurInd}{N - z})}^T \frac{\Dz}{\|\Dz\|_1} \quad \textit{  //as $\CurVar$ is a constant}
\end{aligned}
\end{equation}
where $\Curd$ can be solved as a linear fractional programming (LFP) problem with a closed-form solution.
\autoref{eq:sd-cal} has already considered 
to give a higher degree of approximation for a more expensive feature $j$.  Specifically, when $j$ is ``more expensive'', it means 
$N_j$ is a relatively large number.  A larger $N_j$ will 
lead to a smaller $\frac{I^i_j}{N_j - z_j}$,
giving it a smaller chance to qualify as argmax in \autoref{eq:sd-cal}.
\revision{
With $d^i$ from \autoref{eq:sd-cal}, \RII{Response 2.2}
\name can derive the next approximation plan by  \autoref{eq:policy}
accordingly.
}

\section{Evaluation}

We conducted an evaluation of \name on \NWorkload real inference pipelines sourced from Kaggle and Feast \cite{Feast}, with the aim of demonstrating its ability to reduce inference latency while keeping accuracy loss within acceptable bounds. Our results show that the use of \name leads to a reduction in inference latency of between \MinSpeedup and \MaxSpeedup times compared to the baseline, which involves executing the inference pipeline without any approximation. Moreover, we find that \name can maintain accuracy levels within \MaxAccLoss relative to the baseline.
\revision{We also include RALF \cite{RALF} in the experiments for comparison. \RII{Responses 0.2, 1.2} 
For fair comparison, we give RALF 
an update cost budget no less than the execution time of \name. 
}

\textbf{Workload}.
Despite numerous reports about inference pipelines with expensive aggregations (e.g., \cite{trips-feast,tick-lr,Bearing,tdfraud,student-perf,techton-fraud,demographic,alibaba-fraud}), 
\revision{very few of them have their corresponding real data available as open source.} \RII{Responses 0.1, 1.1}
The ones in FEBench \cite{FEBench} only include feature preparation operators, without any model 
(i.e., no trained model nor training labels provided). 
The ones we used in the evaluation are all publicly available.
Their characteristics are described in \revision{\autoref{tab:workload}}. \RII{Responses 0.1, 1.1}
The pipelines perform regression or classification tasks and employ different numbers of features and models. Certain aggregation queries can generate multiple aggregate features (e.g., in \Trips, the same datastore query produces two features: COUNT and AVERAGE). Each pipeline is also associated with a log of real requests, including information such as user IDs. We execute all the requests and calculate the corresponding averages.  
All inference pipelines were implemented using  Python and Scikit-Learn \cite{sklearn}.

\hspace*{-0.8cm}
\begin{table}[]
\centering
\resizebox{1.1\columnwidth}{!}{
    \hspace*{-1cm}
    \begin{tabular}{|c|c||cccc||cc||c|}
        \hline
        \multirow{3}{*}{\begin{tabular}[c]{@{}c@{}} Pipeline \\(Description) \end{tabular}} &
          \multirow{3}{*}{\begin{tabular}[c]{@{}c@{}} DataSet (Num \\ of Records)\end{tabular}} &
          \multicolumn{4}{c|}{Num of Operators} &
          \multicolumn{2}{c|}{\begin{tabular}[c]{@{}c@{}}Num of\\ Features\end{tabular}} &
          \multirow{3}{*}{\begin{tabular}[c]{@{}c@{}}Num of\\ User\\ Requests\end{tabular}} \\ \cline{3-8}
           &
           &
          \multicolumn{2}{c|}{Datastore} &
          \multicolumn{1}{c|}{\multirow{2}{*}{\begin{tabular}[c]{@{}c@{}}Transfor \\ -mation\end{tabular}}} &
          \multirow{2}{*}{\begin{tabular}[c]{@{}c@{}}Model\\ Inference\end{tabular}} &
          \multicolumn{1}{c|}{\multirow{2}{*}{\textbf{AGG}}} &
          \multirow{2}{*}{\begin{tabular}[c]{@{}c@{}}Non-\\ AGG\end{tabular}} &
           \\ \cline{3-4}
           &
           &
          \multicolumn{1}{c|}{\textbf{AGG}} &
          \multicolumn{1}{c|}{Others} &
          \multicolumn{1}{c|}{} &
           &
          \multicolumn{1}{c|}{} &
           &
           \\ \hline\hline
        \begin{tabular}[c]{@{}c@{}}Trip-Fare \cite{trips-feast}\\ (Predict fare of a trip)\end{tabular} &
          \begin{tabular}[c]{@{}c@{}}NYC \\ Taxi \cite{NYC-Dataset} (3B)\end{tabular} &
          \multicolumn{1}{c|}{2} &
          \multicolumn{1}{c|}{0} &
          \multicolumn{1}{c|}{5} &
          \begin{tabular}[c]{@{}c@{}} LGBM \\ (Regression)\end{tabular} &
          \multicolumn{1}{c|}{3} &
          5 & 1940 \\ \hline
        \begin{tabular}[c]{@{}c@{}}Tick-Price \cite{tick-lr}\\ (Forecast price  of a tick)\end{tabular} &
          \begin{tabular}[c]{@{}c@{}}Forex \\ Tick \cite{Forex-Tick} (1.1B)\end{tabular} &
          \multicolumn{1}{c|}{1} &
          \multicolumn{1}{c|}{6} &
          \multicolumn{1}{c|}{0} &
          \begin{tabular}[c]{@{}c@{}} LR \\ (Regression)\end{tabular} &
          \multicolumn{1}{c|}{1} &
          6 & 4740 \\ \hline
        \begin{tabular}[c]{@{}c@{}} Battery \cite{battery-charge} \\(Predict remaining time \\  to charge battery) \end{tabular} &
          \begin{tabular}[c]{@{}c@{}} NASA Battery \\ \cite{NASA-Battery} (7.3M) \end{tabular} &
          \multicolumn{1}{c|}{5} &
          \multicolumn{1}{c|}{1} &
          \multicolumn{1}{c|}{0} &
          \begin{tabular}[c]{@{}c@{}} LGBM \\ (Regression) \end{tabular} &
          \multicolumn{1}{c|}{{10}} &
          {1} &
          {564} \\ \hline
        \begin{tabular}[c]{@{}c@{}} Turbofan \cite{turbofan-rul} \\ (Predict remaining useful \\ life of turbofan) \end{tabular} &
          \begin{tabular}[c]{@{}c@{}} Turbofan \\ \cite{Turbofan-Dataset} (55M) \end{tabular} &
          \multicolumn{1}{c|}{9} &
          \multicolumn{1}{c|}{0} &
          \multicolumn{1}{c|}{0} &
          \begin{tabular}[c]{@{}c@{}} Random Forest \\ (Regression) \end{tabular} &
          \multicolumn{1}{c|}{9} &
          {0} &
          {769} \\ \hline
        \begin{tabular}[c]{@{}c@{}}Bearing-Imbalance \\ \cite{Bearing}(Detect \\ Imbalance of bearing)\end{tabular} &
          \begin{tabular}[c]{@{}c@{}}Machinery \\ Fault \cite{Machinery-Fault} (95M)\end{tabular} &
          \multicolumn{1}{c|}{8} &
          \multicolumn{1}{c|}{0} &
          \multicolumn{1}{c|}{0} &
          \begin{tabular}[c]{@{}c@{}} MLP \\ (Classification)\end{tabular} &
          \multicolumn{1}{c|}{8} &
          0 &
          338 \\ \hline
        \begin{tabular}[c]{@{}c@{}}Fraud-Detection \cite{tdfraud}\\ (Detect fraudulent click \\ of user)\end{tabular} &
          \begin{tabular}[c]{@{}c@{}}TalkingData Click\\  \cite{TalkingData-Clicklog} (242M)\end{tabular} &
          \multicolumn{1}{c|}{3} &
          \multicolumn{1}{c|}{0} &
          \multicolumn{1}{c|}{6} &
          \begin{tabular}[c]{@{}c@{}} XGB \\ (Classification)\end{tabular} &
          \multicolumn{1}{c|}{3} &
          6 & 8603 \\ \hline
        \begin{tabular}[c]{@{}c@{}}Student-QA \cite{qa-correctness}\\ (Predict correctness \\ of a question ) \end{tabular} &
          \begin{tabular}[c]{@{}c@{}} Game Log \\ \cite{Student-GameLog} (26M) \end{tabular} &
          \multicolumn{1}{c|}{13} &
          \multicolumn{1}{c|}{0} &
          \multicolumn{1}{c|}{0} &
          \begin{tabular}[c]{@{}c@{}} Random Forest \\ (Classification) \end{tabular} &
          \multicolumn{1}{c|}{21} &
          {0} &
          {471} \\ \hline
    \end{tabular}
}
\caption{\revision{Real Inference Pipelines}}
\label{tab:workload}
\vspace{-3em}
\end{table}

\vspace{-1em}
\textbf{System Setup}. \name is implemented using Python.
We used ClickHouse \cite{clickhouse}, an open-source OLAP DBMS designed for real-time data analytics with support for online sampling, as our datastore.
Nonetheless, it is worth noting that \name is not tied to any specific data store solution and may be used with other databases, such as MySQL, or data analytics frameworks, such as Pandas and Dask, without losing its advantages.
All the experiments were run on servers with Intel Xeon E5-2620 CPU (2.1 GHz with 8 physical cores), 256 GB of memory, and 745 GB of Intel DC S3610 Series SSD.

\textbf{Metrics}.
To provide a comprehensive evaluation of \name, we 
run the experiments five times and report the average latency of the system when serving all real user requests, as well as its speedup compared to the baseline. Additionally, we report the 
accuracy.  \revision{Unless stated otherwise,
\RII{Response 1.1}
the accuracy is measured using the 
true label in the hold-out set}.
We use \ScoreCls and \ScoreReg to measure the accuracy of classification and regression pipelines, 
respectively.

\textbf{Default Configuration}.
During the evaluation process, we have employed a default configuration that is shared by all workloads for \name. Specifically, we set the sampling ratio for the initial plan as $\InitRatio=\DefaultInitRatio$.
Following typical online aggregation systems \cite{GOLA}, we set the step size $\gamma$ 
as 1\% of the total number of records 
across all features.
The confidence level 
$\ConfLevel$ 
is set to $\DefaultConfLevel$. 
For classification tasks, we set the error bound $\ErrorBound=0$ to ensure precise results. For the regression tasks, we set $\ErrorBound=MAE$, where $MAE$ is the mean absolute error of the pre-trained model in the test set. 
Additionally, we set $m=\DefaultM$ as the number of samples for QMC.

\subsection{End to End Performance}\label{sec:e2e}
\begin{figure}[t]
    \centering
    \begin{subfigure}{\linewidth}
        \includegraphics[width=\linewidth]{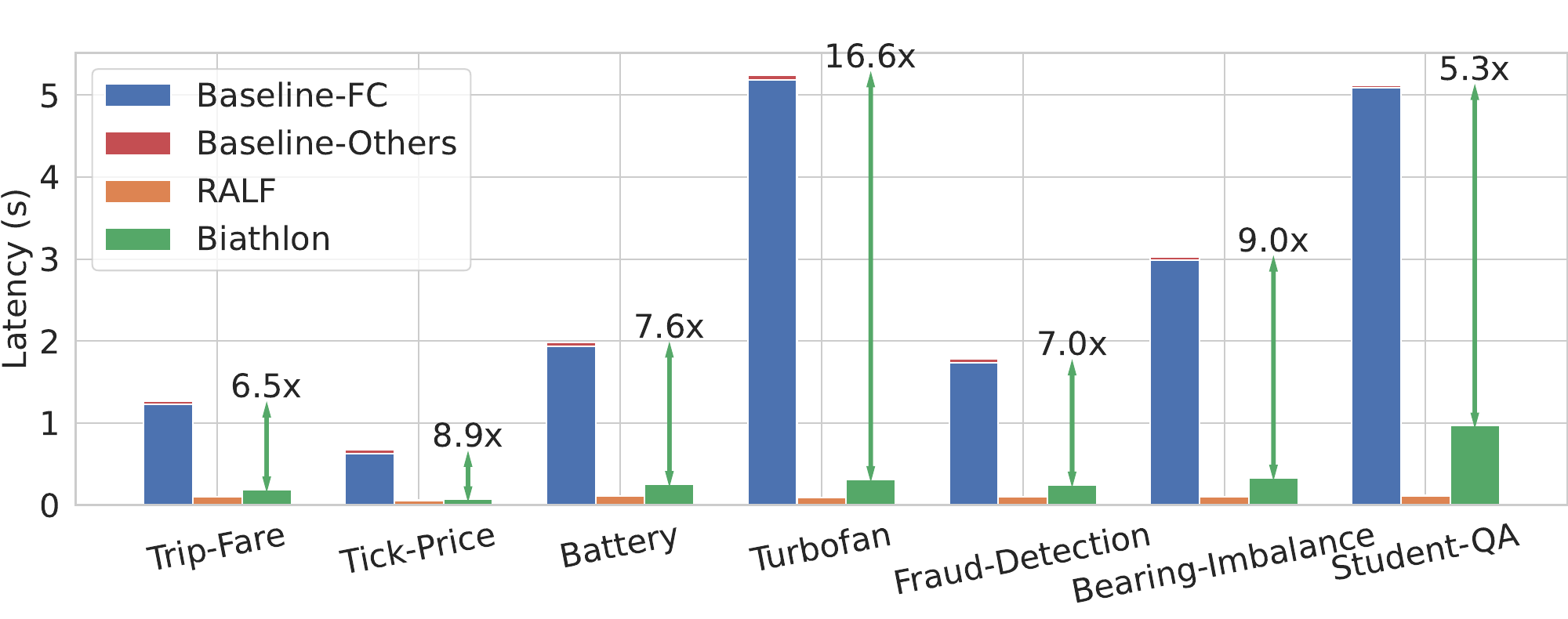}
    \end{subfigure} \hfill
    \begin{subfigure}{\linewidth}
        \includegraphics[width=\linewidth]{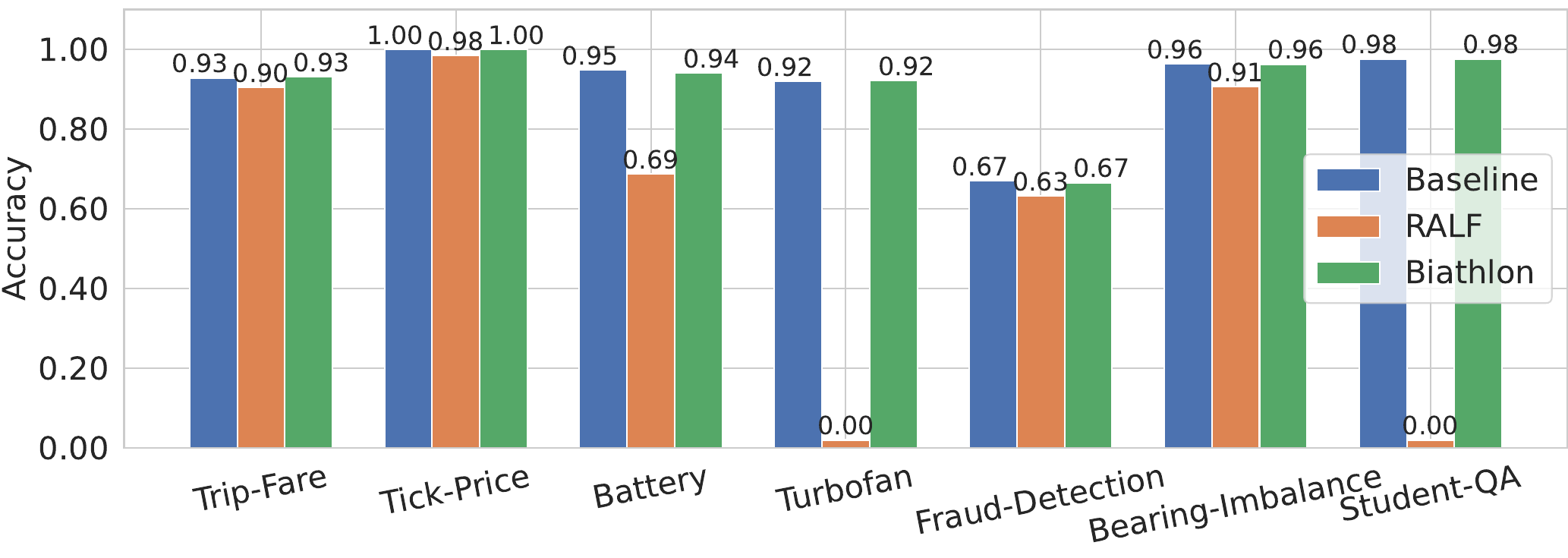}
    \end{subfigure} \hfill
    \vspace{-2em}
    \caption{
    Latency and Accuracy (default configuration)} 
    \label{fig:e2e-default}
\end{figure}

Figure \ref{fig:e2e-default} presents the performance evaluation of \name on the \NWorkload inference pipelines under the default configuration. 
The top figure illustrates the latency comparison 
among the baseline, \revision{RALF}, \RII{Responses 0.2, 1.2} and \name on the \NWorkload workloads. It is evident that feature computation (FC) is the most time-consuming and dominates the latency of the baseline.  The baseline incurs a latency of
more than a second on most pipelines, which is not ideal for user-facing applications.
Despite their distinct characteristics, running the inference pipelines on \name shows a significant speedup, ranging from \MinSpeedup to \MaxSpeedup. More importantly, \name is able to achieve sub-second real-time response latency on all pipelines.
The bottom figure shows the accuracy of each workload in \name. 
It is evident that \name achieves its real-time latency
with almost no accuracy loss with respect to 
the exact baseline.

\revision{RALF, in contrast, 
despite having very low latency,
indeed suffers from accuracy loss and unbounded error.
Specifically, as a feature store, RALF generally 
exhibits lower accuracy in pipelines with frequent updates (Tick-Price).
In pipelines with slow error feedback 
(Trip-Fare and Fraud-Detection),
\RII{Responses 0.2, 1.2}
RALF also demonstrates lower accuracy 
due to the inability to update its feedback loop promptly.
Furthermore, in pipelines with many new and unseen items
(Battery, TurboFan, Bearing Imbalance and Student-QA), 
RALF has even poorer accuracy
because it would never compute the feature value online. 
Instead, for any compulsory cache miss \cite{hennessyComputerArchitectureFifth2011}
(i.e., item that has never been in the cache), 
RALF would assume a default value for that feature 
and rely solely on the error feedback loop to select those items for pre-computation in the future. Unfortunately, these pre-computed items are seldom seen again in subsequent requests in those workloads.
}

\begin{figure}
\centering
  \centering
  \includegraphics[width=0.5\linewidth]{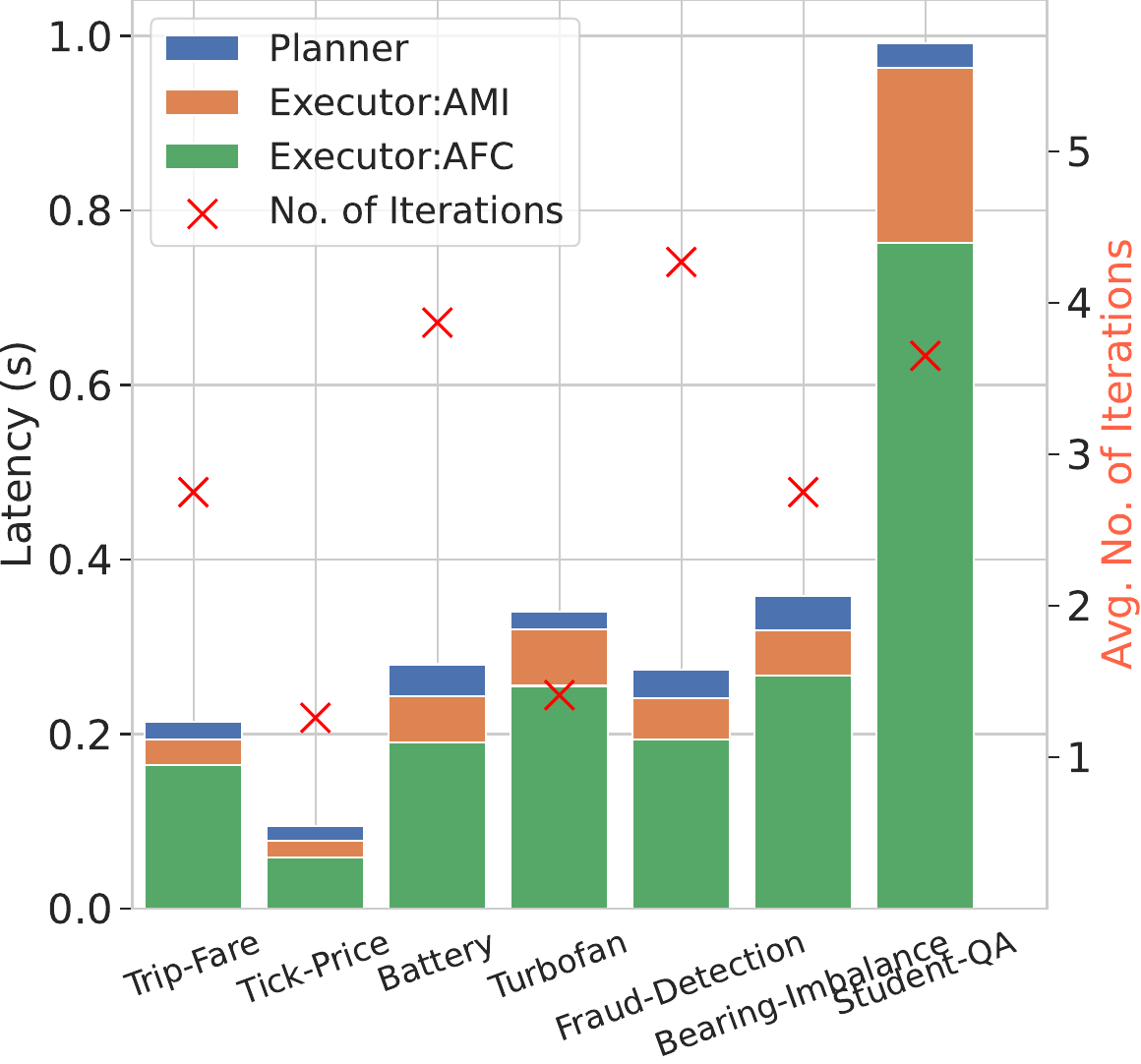}
  \vspace{-0.2cm}
  \captionof{figure}{Latency Breakdown of \name}
  \label{fig:lat-bd}
\vspace{-1em}
\end{figure}

\begin{figure*}
    \centering
    \includegraphics[width=\linewidth]{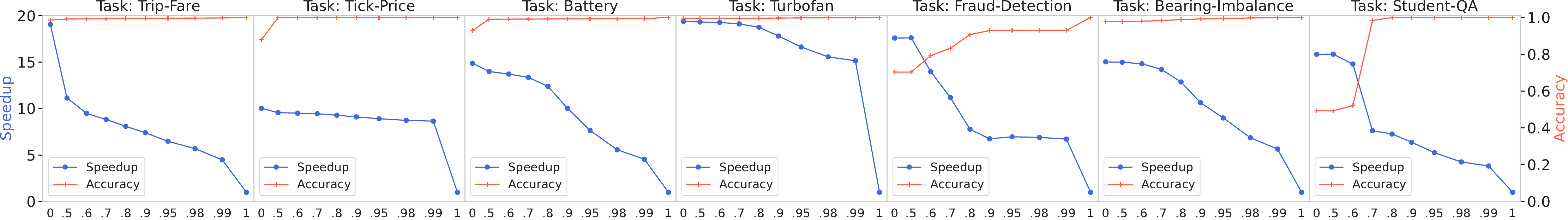}
    \caption{Varying Confidence Level $\ConfLevel$}
    \label{fig:vary_conf}
\end{figure*}

Figure \ref{fig:lat-bd} shows a breakdown of the latency components for each workload in \name. 
The breakdown comprises three parts: AFC (which measures the cost of feature preparation), AMI (which considers the cost of model inference and the overhead of QMC in
estimating the inference result uncertainty $U_y$),
and Planner (which \revision{online} \RII{Response 2.2} devises the new plan based on the inference variance reduction via the computation of the Main Effect Indices of individual features).

Within \name, 
the majority of latency is still dominated by approximated feature computation (AFC), which involves I/O. However, that time has been significantly reduced when compared with the baseline because only a small fraction of data
(about 5.4\% to 14.4\% according to our profiling) is actually touched. 
\revision{Specifically, the average numbers of iterations 
consumed by each pipeline, as shown in Figure \ref{fig:lat-bd},
\RII{Response 2.3} are all less than 5, 
indicating all pipelines are able to satisfy \autoref{eq:accuracy_target} and 
early stop.}
\revision{Furthermore, 
we also empirically measure the percentage of inference requests whose
actual error, i.e., $|Y - \hat{y}|$, falls within
our given default error bound $\ErrorBound$
($\ErrorBound=0$ for classification,
$\ErrorBound=MAE$ for regression).  \RII{Response 2.4}
We found that all pipelines have 95\% to 100\% of 
their inference requests 
with real errors that fall within the required error bound, perfectly aligning with the specified confidence level $\ConfLevel = \DefaultConfLevel$.
}

\subsection{Varying the confidence level $\ConfLevel$}
\label{sec:varyconf}
In this experiment, we aim to study the impact of 
the confidence level $\ConfLevel$ 
in \autoref{eq:accuracy_target}.
\revision{
Since Equation \ref{eq:accuracy_target} 
defines the guarantee 
based on the 
error between 
the prediction made by the baseline
and the prediction made by \name, 
here we calculate the accuracy \RII{Response 1.1} 
based on using the exact value predicted by the baseline
as the oracle label. 
}
Figure \ref{fig:vary_conf} shows the result
of varying $\ConfLevel$.
It can be observed that the speedup of \name decreases as the required confidence level $\ConfLevel$ increases. This is expected since \name needs to retrieve more data to achieve a higher level of required confidence. When a confidence level of 1.0 is required, \name necessitates exact features as input and does not provide any speedup. However, apart from that, \name maintains a substantial speedup even when the required confidence level is as high as 0.99. 
Some pipelines maintain near-perfect accuracy while achieving speedup, irrespective of the confidence level value. 
For these pipelines, the initial approximation plan produces feature computations that yield sufficiently accurate inference results for the majority of requests. Indeed, there are still improvements in accuracy
with higher confidence levels but they are not readily apparent in the figures.
For instance, in the Turbofan pipeline, the $r^2$-score escalates from 0.9943 to 0.9982 as the confidence level rises from 0.0 to 0.99.

\subsection{Varying the error bound $\ErrorBound$}

\begin{figure}
    \includegraphics[width=\linewidth]{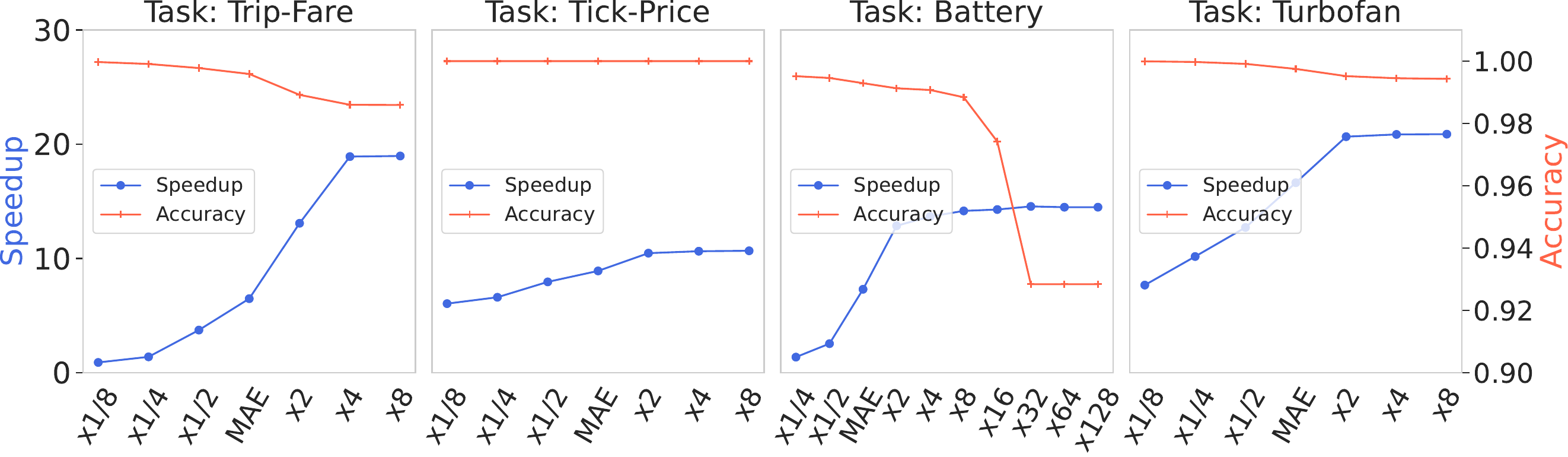}
    \vspace{-1.5em}
    \caption{Varying Error Bound $\ErrorBound$ (Regression Only). } 
    \label{fig:vary_error}
\end{figure}

In this experiment, we aim to study the impact of 
the error bound $\ErrorBound$ 
in \autoref{eq:accuracy_target}.
Figure \ref{fig:vary_error} presents the results of speedup and accuracy in relation to varying error bound values $\ErrorBound$.
\RII{Response 1.1}\revision{Same as the above,
the accuracy of \name is 
calculated using 
the exact value predicted by the baseline as the oracle label.}
Only the results of regression pipelines are shown since the others involve classification, which cannot tolerate any error.

From the figure, we can observe that the speedup of \name increases as the tolerable error $\ErrorBound$ increases. This is expected because a higher value of $\ErrorBound$ allows \name to satisfy \autoref{eq:accuracy_target} more easily, resulting in fewer data being retrieved. 
As the error bound $\ErrorBound$ continues to increase, the speedup eventually 
remains stable. This is because Biathlon can then easily satisfy \autoref{eq:accuracy_target} after the first iteration when given a very large $\ErrorBound$. Subsequently, further increasing $\ErrorBound$ 
would not reduce the number of iterations any further.

Increasing the error bound would naturally have a negative effect on accuracy because more results with larger errors can satisfy \autoref{eq:accuracy_target}. The Tick-Price pipeline is insensitive to this effect because the samples drawn by Biathlon in the first iteration already provide more than enough samples even for the tightest error bound. Therefore, the relaxation of the error is immaterial for this pipeline.

\section{Related Work}

Many ML serving systems have been proposed to enhance the execution efficiency of inference pipelines. Some systems focus on simplifying the deployment process through containerized execution \cite{Clipper, Rafiki} and in-application execution \cite{MLwithMLNet, MSCloud2Edge}, while others aim to accelerate model inference via resource-sharing \cite{Pretzel}, compilation \cite{TVM,Wield}, scheduling \cite{Pretzel,Wu2022ServingAO}, and caching \cite{Pretzel}. 
Some systems \cite{Raven, RavenInitial, hummingbird, HeDongTQP, ContainerizedUDF} propose integrating machine learning and database into unified frameworks for inference pipelines, thereby facilitating cross-optimization \cite{Raven} between feature computation operators and inference operators. 
However, these approaches primarily focus on enabling or optimizing ML inference without leveraging the resilience intrinsic to machine learning models, like how \name did. Some recent systems \cite{Willump, ACCMPEG, LASER} also leverage some other statistical properties besides resilience to expedite inference pipelines, but are limited to Linear models \cite{LASER} or lack accuracy guarantee \cite{Willump,ACCMPEG}. 
\revision{In general, we do not recommend using Biathlon for deep model pipelines.
Deep learning models tend to be computationally expensive compared to traditional non-deep models. Since Biathlon conducts multiple model inferences for each inference request during quasi-Monte Carlo (QMC), the resulting overhead can outweigh the benefits gained from feature approximation. \RII{Response 2.5}
However, it is worth noting that accelerating deep learning pipelines is an important area of research and there are corresponding solutions available \cite{NoScope, Everest, yan2024decoding}.}

\section{Conclusion and Future work}
This paper presents \name, an innovative ML serving system specifically tailored for data science and industry inference pipelines. Developed to address the stringent user-facing latency demands of real-time inference, \name incorporates several key components: approximate query processing from the database area to compute feature approximately, uncertainty propagation from statistical analysis to estimate inference uncertainty, feature importance based on Sobol Indices from model interpretability to assess the contribution of individual features to the inference uncertainty, and an iterative optimization algorithm for determining the best approximation plan. 

\name offers maximum speedup with a probabilistic guarantee of bounded error and achieves a speedup ranging from \MinSpeedup to \MaxSpeedup on real pipelines without a noticeable loss in accuracy. 
\revision{Inherited from online aggregation, there are operators that \name does not approximate (e.g., Top-K).
We believe that, for such cases,  \RII{Response 3.1}
the feature store approach, as demonstrated by RALF, can be a viable alternative. However, it is essential to remark that the feature store approach (including RALF) also has its limitations, such as the absence of error bounds or being restricted to a limited set of workloads. Therefore, we believe that the feature store caching methodology of RALF and our AQP approach can complement each other, opening up an intriguing avenue for future research.
}

\section*{Acknowledgement}
This work is partially supported Hong Kong General Research Fund (14208023), Hong Kong AoE/P-404/18, and the Center for Perceptual and Interactive Intelligence (CPII) Ltd under InnoHK supported by the Innovation and Technology Commission.
We also thank Professor Yufei Tao for his insightful comment about this work. 

\bibliographystyle{ACM-Reference-Format}
\bibliography{main}

\clearpage
\appendix

\begin{table}[h]
\begin{tabular}{|l|l|}
\hline
Symbol         & Meaning                                                                                                                            \\ \hline
$\delta, \tau$ & error bound, confidence level                                                                                                      \\ \hline
$z^i_{1:k}$    & \begin{tabular}[c]{@{}l@{}}approximation plan at $i$-th iteration, \\ where $z_j^i$ is the sample size of feature $j$\end{tabular} \\ \hline
$N_j$          & max. number of selectable records of feature $j$                                                                                   \\ \hline
$C^z$          & cost of the approximation plan $z$                                                                                                 \\ \hline
$\hat x_{1:k}$ & approximated feature values computed with samples                                                                                                        \\ \hline
$X$            & exact feature values computed with ALL data                                                                                                       \\ \hline
$\hat y$       & approximate inference result computed from $\hat x$                                                                                                \\ \hline
$Y$            & exact inference result computed from $X$                                                                                                         \\ \hline
$U_x$          & estimated uncertainty of approximated features                                                                      \\ \hline
$U_y$          & estimated uncertainty of inference result                                                                      \\ \hline
$I_j^i$        & main effect index for feature $j$ at $i$-th iteration                                                                              \\ \hline
$Var(Y|z^{i})$     & \begin{tabular}[c]{@{}l@{}}variance of inference result with the approximate \\ plan at $i$-th iteration \end{tabular} \\ \hline

$d^i$          & \begin{tabular}[c]{@{}l@{}}direction characterized by the maximum reduction \\in inference uncertainty at $i$-th iteration\end{tabular} \\ \hline

$\InitRatio$   & sampling ratio of the initial approximation plan  \\ \hline

$\StepSize$    & step size for every iteration  \\ \hline

\end{tabular}
\caption{
\revision{Notation Table}}
\label{tab:notation}
\vspace{-2em}
\end{table}

\section{Varying the initial sampling ratio $\InitRatio$ } 
\begin{figure*}
\centering
\begin{minipage}{\linewidth}
  \centering
  \includegraphics[width=\linewidth]{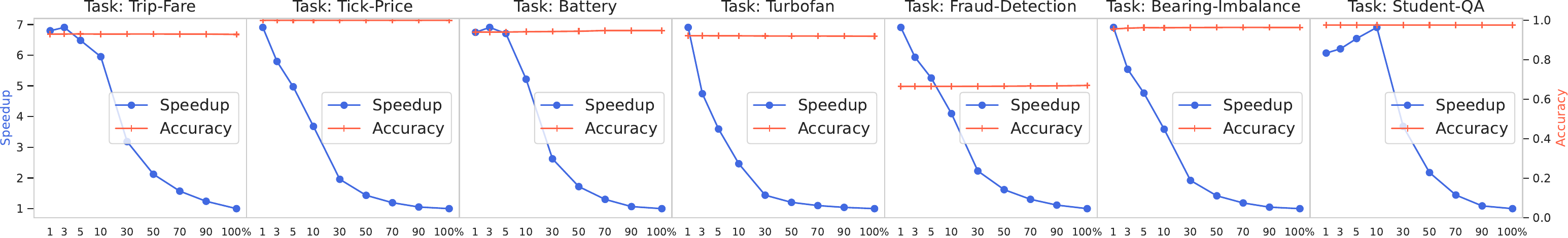}
  \vspace{-2em}
  \captionof{figure}{Varying Initial Sampling Ratio $\InitRatio$}
  \label{fig:vary_init}
\end{minipage}
\begin{minipage}{\linewidth}
  \centering
  \includegraphics[width=\linewidth]{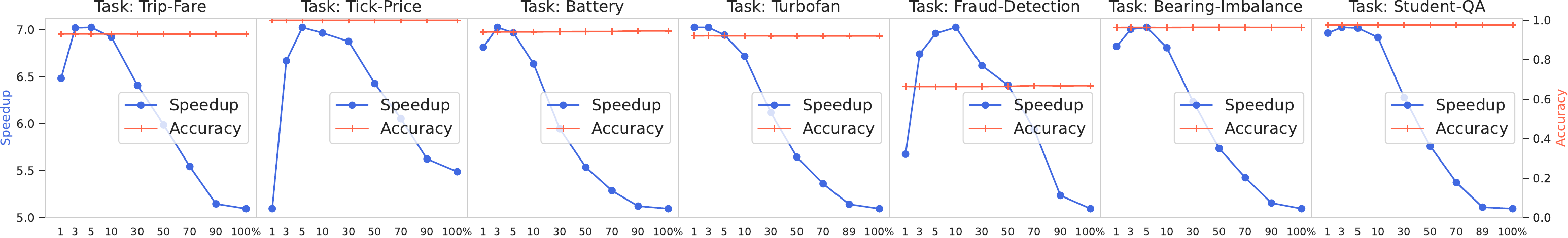}
  \vspace{-2em}
  \captionof{figure}{Varying Step Size $\gamma$}
  \label{fig:vary_step}
\end{minipage}
\end{figure*}
In this experiment, we aim to study the impact of 
the initial sampling ratio $\InitRatio$ 
with respect to the inference pipelines.
 The results are depicted in Figure \ref{fig:vary_init}. 
 The observed decrease in speedup with excessively high or low values of $\InitRatio$ can be attributed to specific factors. When $\InitRatio$ is set too high, the initial approximation plan may allocate an excessive number of samples, potentially overshooting the requirement to satisfy Equation \ref{eq:accuracy_target}. Conversely, setting $\InitRatio$ too low may result in Biathlon requiring more iterations to return, thus increasing the associated overhead (e.g. QMC computations).
 As shown in Figure \ref{fig:vary_init}, there is often a sweet spot for $\InitRatio$ that achieves the highest speedup, and the location of this sweet spot depends on the individual pipelines. Tuning hyperparameters like this for each pipeline is feasible and can be done offline, but it falls outside the current scope of \name and would be a topic for future work. 
 
 It is worth noting that the value of $\InitRatio$ has less influence on the final accuracy since all pipelines in this experiment have a probability of at least 0.95 to have an error of 0 or below the mean absolute error.

\section{Varying the step size $\StepSize$} 

This experiment aims to investigate the impact of the step size $\StepSize$ on the inference pipelines. To recap, a step size that is too large would overshoot the number of samples required to meet \autoref{eq:accuracy_target}. Conversely, a step size that is too small would increase the number of iterations and the associated overheads (e.g., QMC). As shown in Figure \ref{fig:vary_step}, there is often a sweet spot for $\StepSize$ that achieves the highest speedup, and the location of this sweet spot depends on the individual pipelines. Tuning hyperparameters like this for each pipeline is feasible and can be done offline, but it also regards that
as the future work of \name.

Similar to $\InitRatio$, the value of $\StepSize$ has less influence on the final
accuracy since all pipelines in this experiment have a probability of
at least 0.95 to have an error of 0 or below the mean absolute error.

\section{Varying the number of approximated aggregation operators}
\begin{figure}
  \centering
  \includegraphics[width=0.5\linewidth]{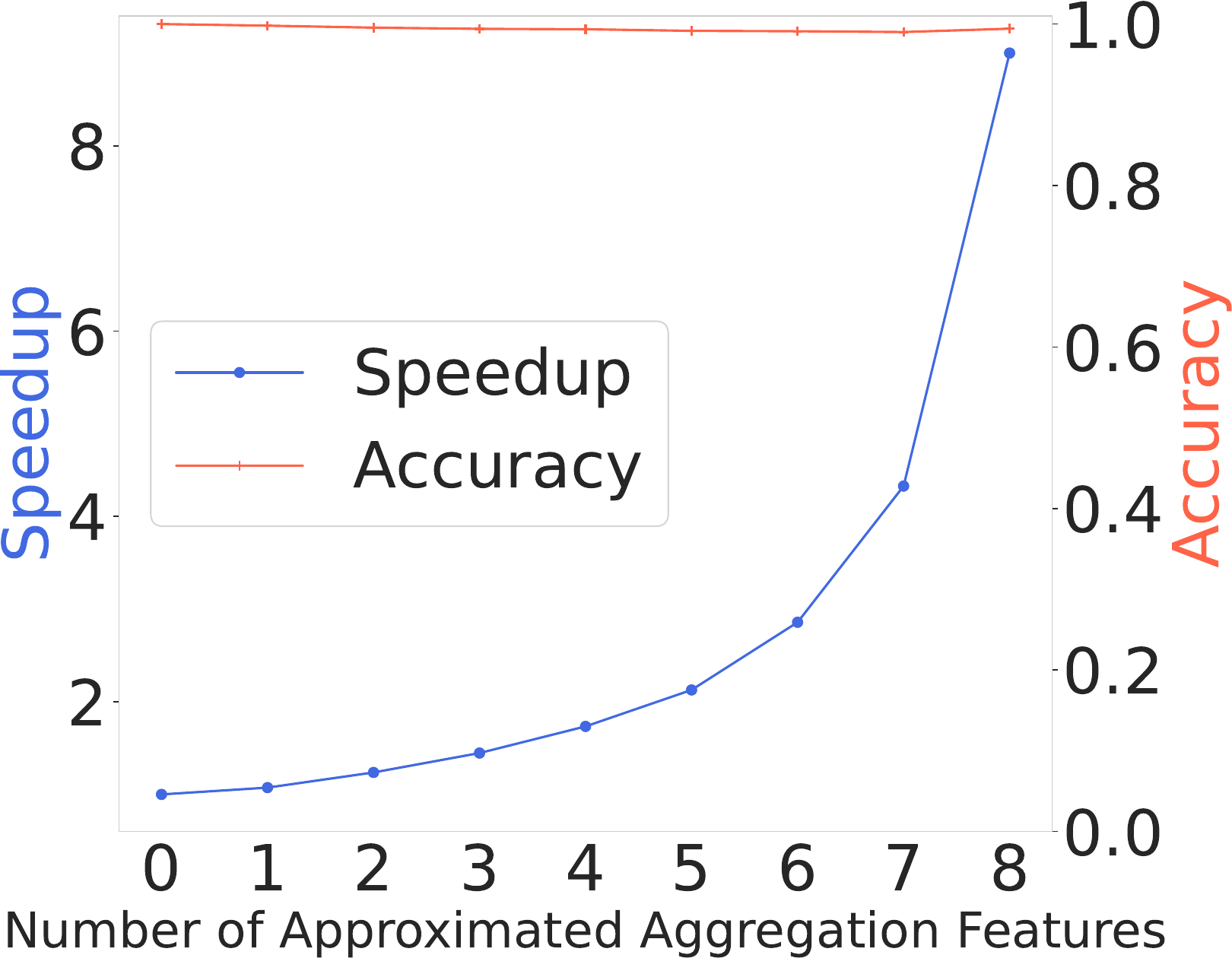}
  \captionof{figure}{Varying Number of Approximated Operators}
  \label{fig:vary_nagg}
\end{figure}
In this experiment, we aim to study the impact of the number of approximated aggregation operators in a pipeline. We vary the number of aggregation features in the Bearing-Imbalance that undergo approximation. 
We chose the Bearing-Imbalance pipeline because it has eight aggregation features, allowing us to control the number of features being approximated from 0 (i.e., all features are exact) to 8 (i.e., all features are subjected to \name for approximation).

Figure \ref{fig:vary_nagg} shows the results, where we can observe that the speedup of the pipeline increases with the number of aggregation features being approximated by \name. Since different aggregation features contribute different shares of importance to the final prediction accuracy, the speedup improvement is not necessarily linear. The accuracy of the final prediction remains high and stable, indicating that \name fully leverages the model resilience to accelerate without jeopardizing the inference quality.

\section{Experiment on MEDIAN} 
\begin{figure*}[h!]
    \centering
    \includegraphics[width=\linewidth]{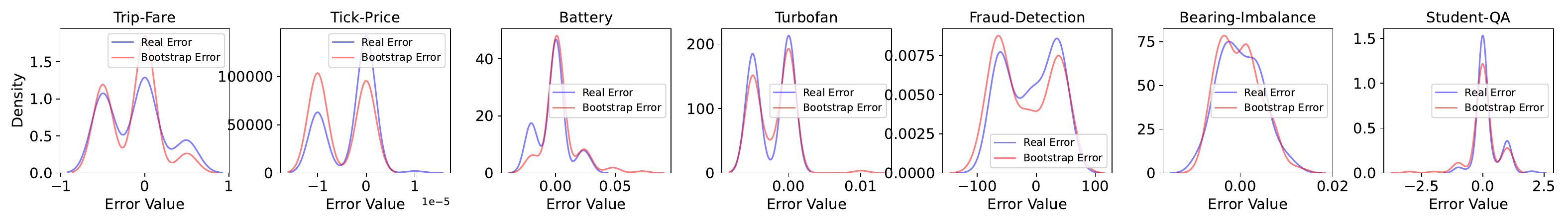}
    \vspace{-3em}
    \caption{Example Error Distribution of Median Feature}
    \label{fig:2.1.2}
\end{figure*}

\begin{figure*}
\begin{minipage}{0.135\linewidth}
\centering
\includegraphics[width=\linewidth]
{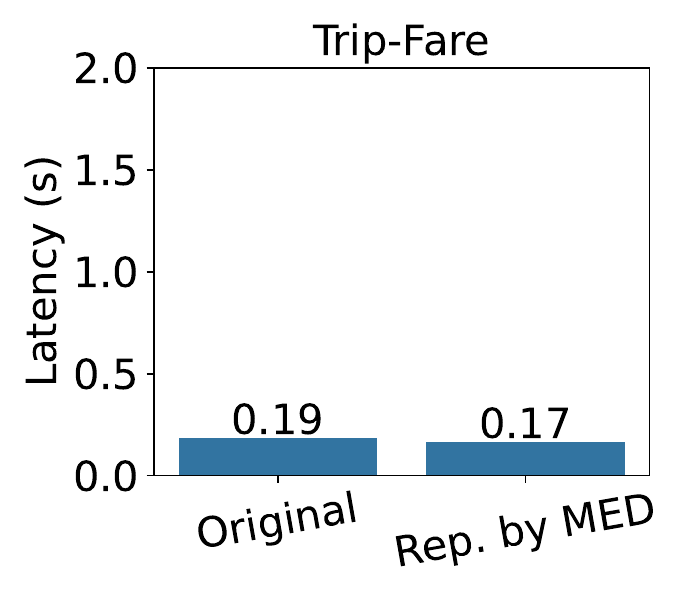}
\end{minipage}
\begin{minipage}{0.135\linewidth}
\centering
\includegraphics[width=\linewidth]{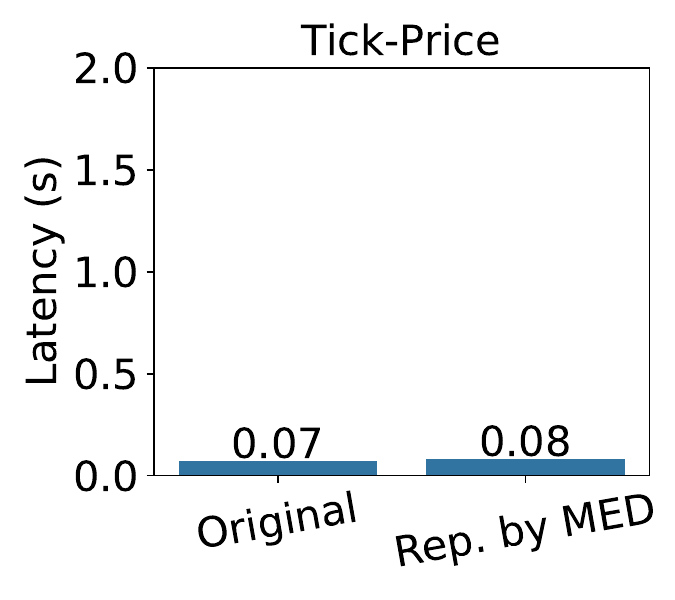}
\end{minipage}
\begin{minipage}{0.135\linewidth}
\centering
\includegraphics[width=\linewidth]{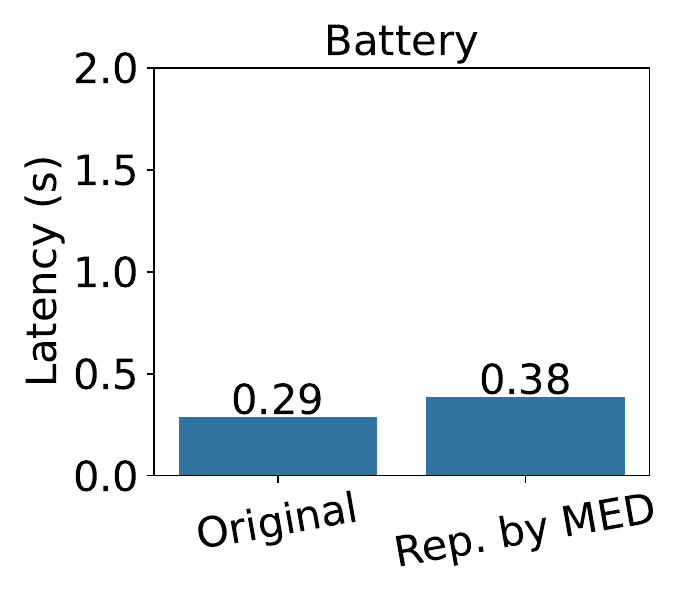}
\end{minipage}
\begin{minipage}{0.135\linewidth}
\centering
\includegraphics[width=\linewidth]{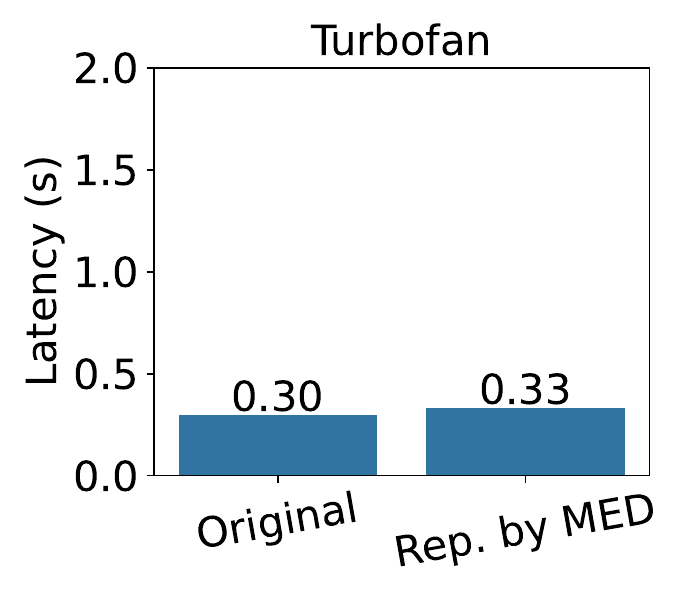}
\end{minipage}
\begin{minipage}{0.135\linewidth}
\centering
\includegraphics[width=\linewidth]{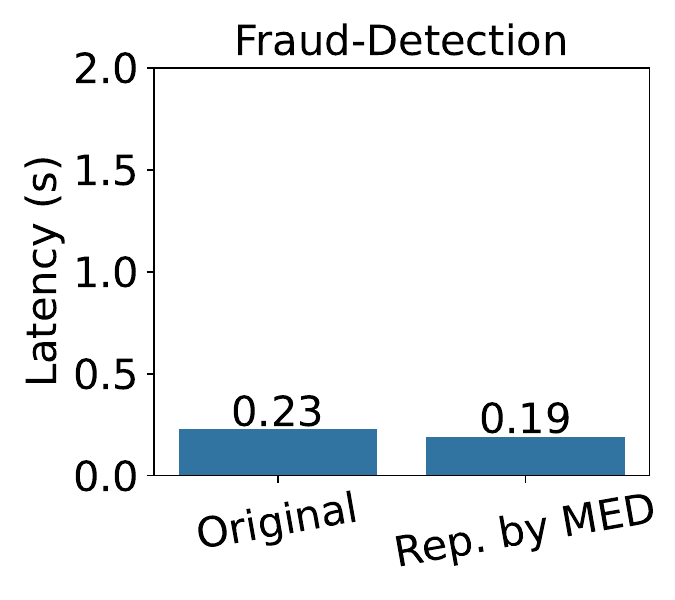}
\end{minipage}
\begin{minipage}{0.135\linewidth}
\centering
\includegraphics[width=\linewidth]{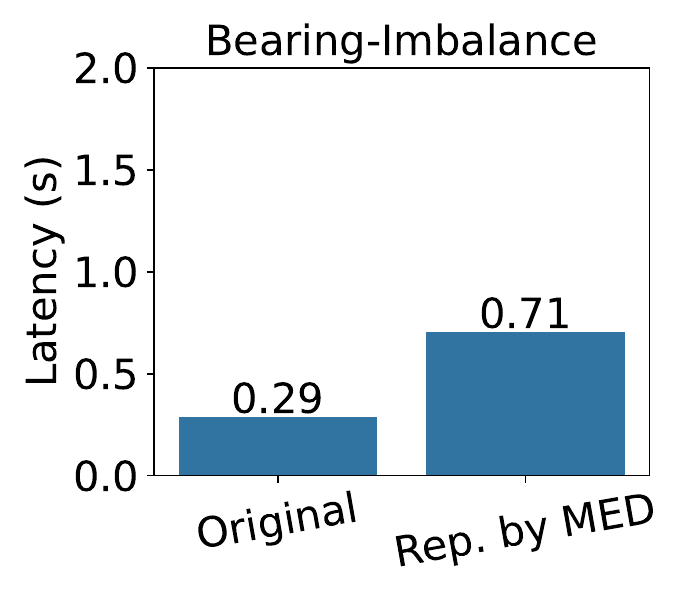}
\end{minipage}
\begin{minipage}{0.135\linewidth}
\centering
\includegraphics[width=\linewidth]{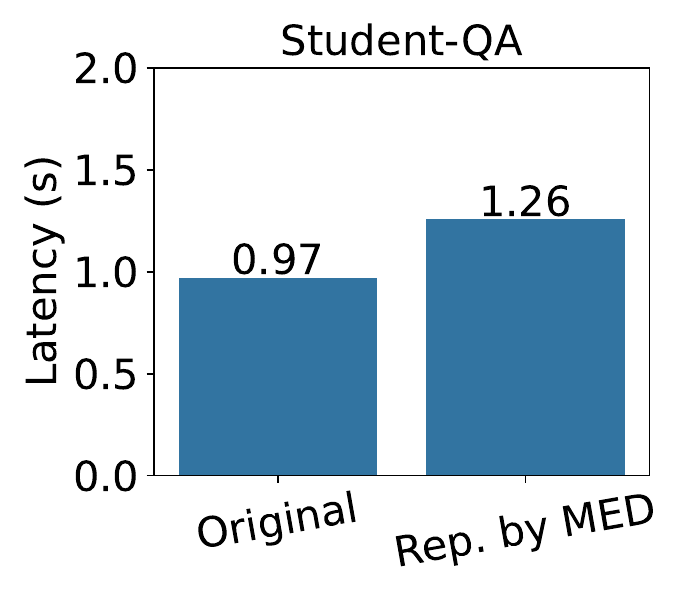}
\end{minipage}
\begin{minipage}{0.135\linewidth}
\centering
\includegraphics[width=\linewidth]
{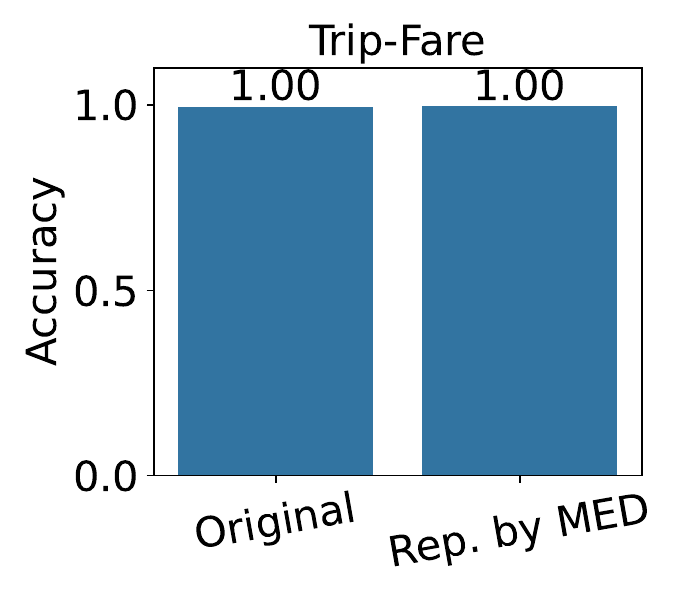}
\end{minipage}
\begin{minipage}{0.135\linewidth}
\centering
\includegraphics[width=\linewidth]{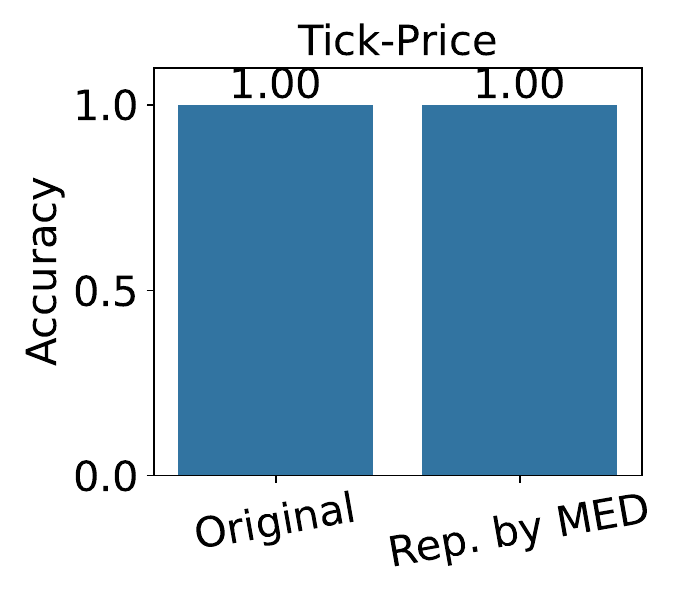}
\end{minipage}
\begin{minipage}{0.135\linewidth}
\centering
\includegraphics[width=\linewidth]{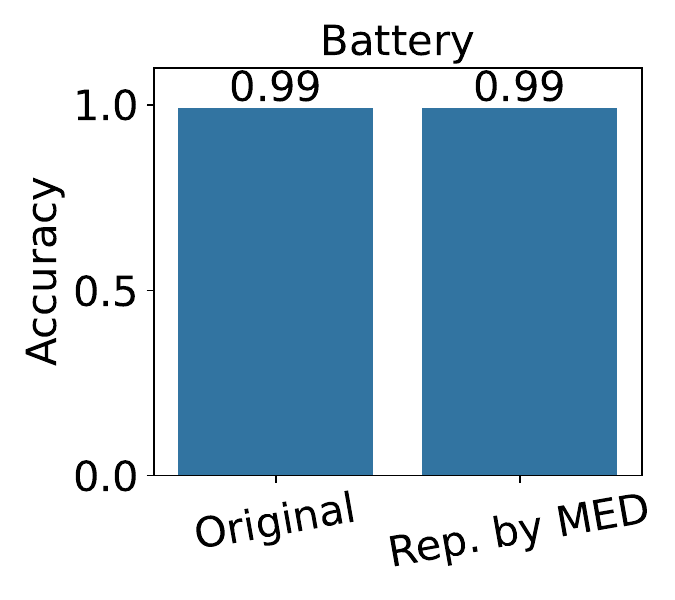}
\end{minipage}
\begin{minipage}{0.135\linewidth}
\centering
\includegraphics[width=\linewidth]{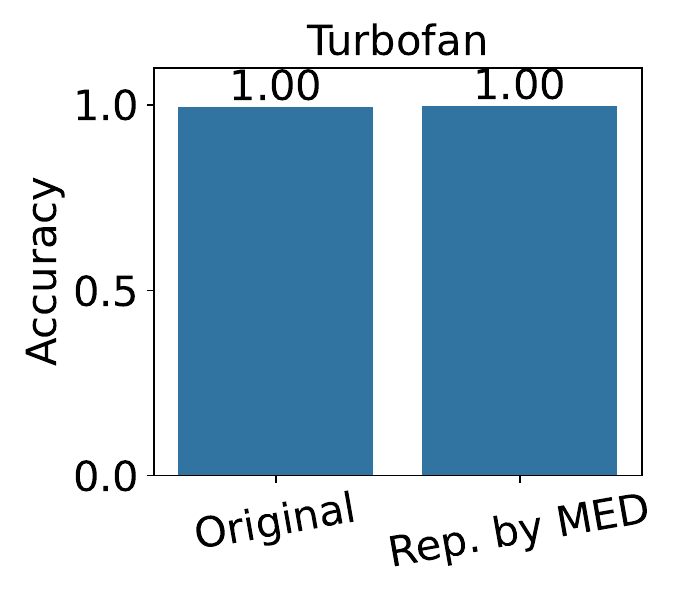}
\end{minipage}
\begin{minipage}{0.135\linewidth}
\centering
\includegraphics[width=\linewidth]{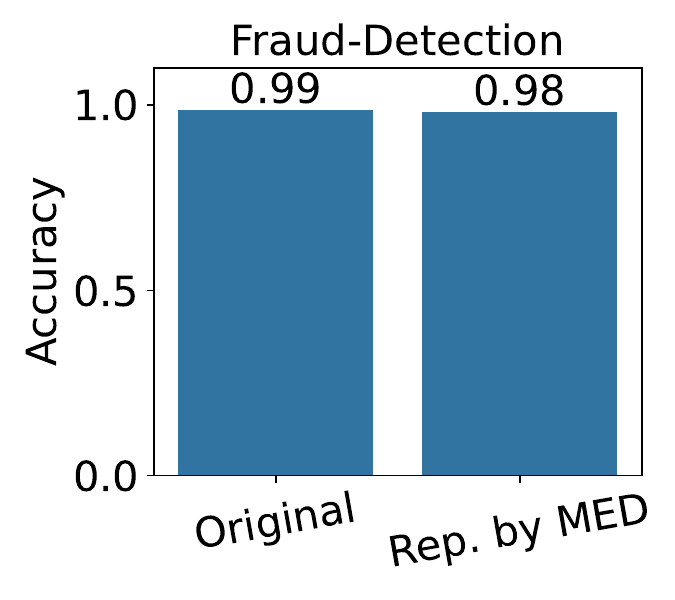}
\end{minipage}
\begin{minipage}{0.135\linewidth}
\centering
\includegraphics[width=\linewidth]{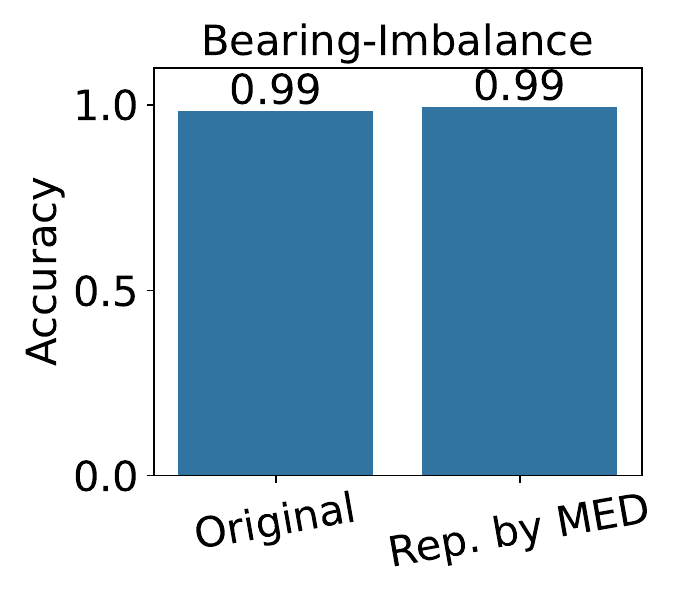}
\end{minipage}
\begin{minipage}{0.135\linewidth}
\centering
\includegraphics[width=\linewidth]{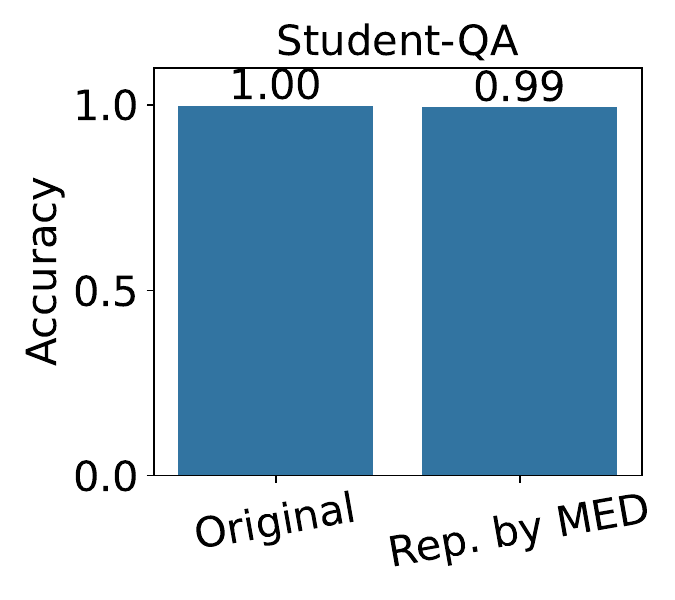}
\end{minipage}
\vspace{-1em}
\caption{Original pipelines vs Pipelines with operators substituted by MEDIAN} 
\label{fig:2.1.3}
\end{figure*}

\begin{figure*}
    \centering
    \includegraphics[scale=0.5]{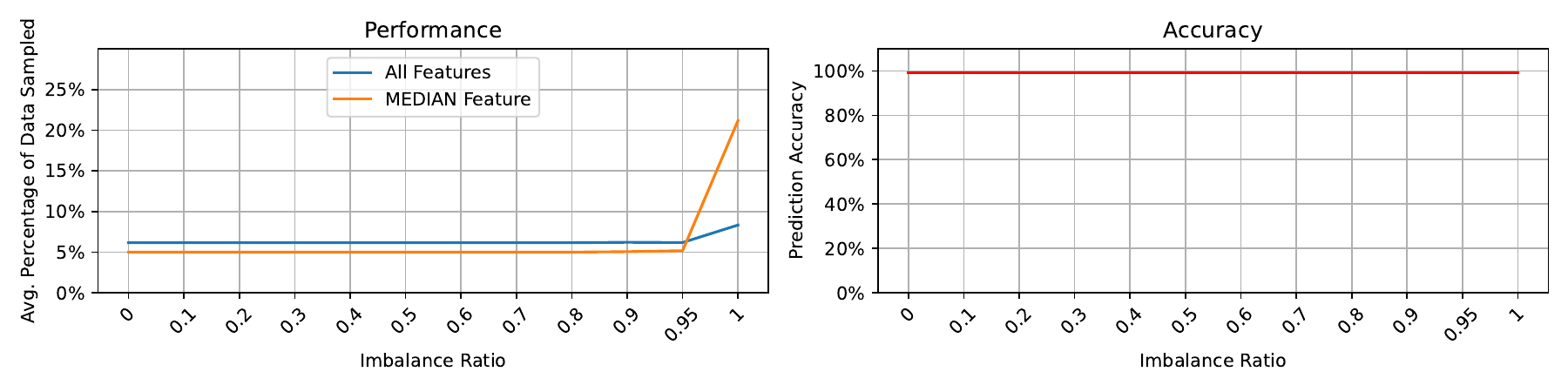}
    \vspace{-2em}
    \caption{Varying imbalance ratio for the median feature in Bearing-Imbalance pipeline.  }
    \label{fig:3.2.1}
\end{figure*}

\begin{figure*}
    \centering
    \includegraphics[scale=0.5]{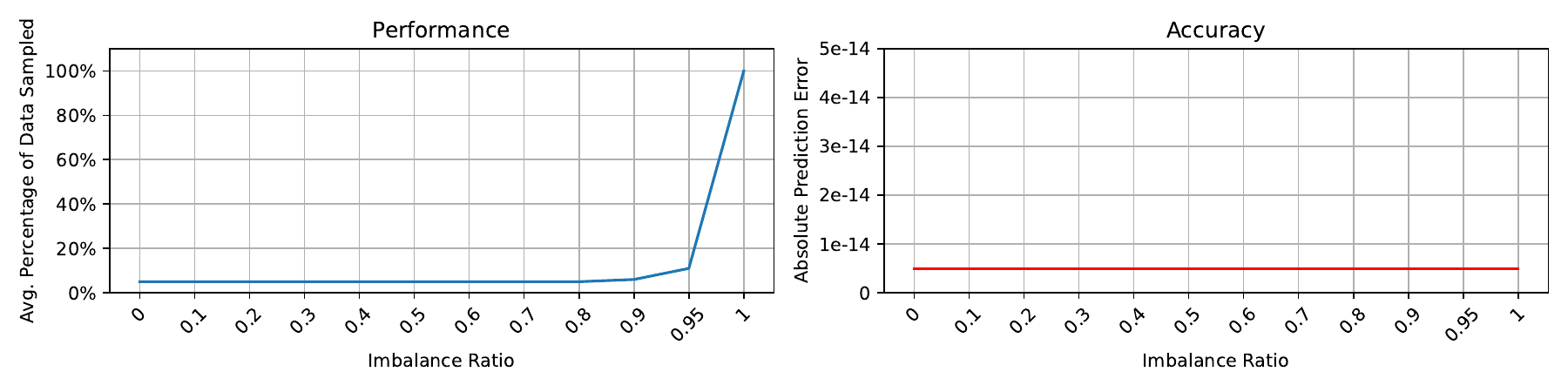}
    \vspace{-2em}
    \caption{Varying imbalance ratio for the median feature in Tick-Price pipeline}
    \label{fig:3.2.2}
\end{figure*}

\RII{Response 2.1}
\revision{
For holistic aggregations MEDIAN and QUANTILE, whose error distribution is unknown, we use Empirical Bootstrap \cite{Bootstrap} to obtain their error distributions. Empirical Bootstrap is a non-parametric method. 
Since our original real pipelines do not use MEDIAN and QUANTILE,
we conducted a new set of experiments where
we replaced all the AVERAGE operators in the pipelines by MEDIAN. (In the Fraud-Detection pipeline, we substituted the COUNT operators by MEDIAN instead, as 
it has no AVERAGE operators).
Since now some features have changed, we re-trained the models of those pipeline accordingly.

Figure \ref{fig:2.1.2} shows one instance of error distribution of the MEDIAN feature for each modified pipeline. 
We can see that with MEDIAN, although the error distribution is unknown,
bootstrapping can indeed capture them fairly accurately.

Figure \ref{fig:2.1.3} shows the performance of Biathlon 
on the original pipelines and their counterparts after 
being replaced by MEDIAN. Since the models have changed, their input resilience has also changed, which explains the differences in inference latency and accuracy.
Nonetheless, we can see that Biathlon remains good in latency and accuracy.
The Student-QA pipeline has an average latency
slightly over a second 
because this pipeline originally has the highest number of 
features where 8 of them are computed by AVG.
Now it has 8 MEDIAN operators, 
which incurs more overhead because of bootstrapping.
Nonetheless, recall that the results here are based on the default 
configurations (e.g, all pipelines use the same step size).
In practice, we can fine-tune the configuration for each pipeline to achieve even better latency. 
For this pipeline, after we manually tuned the initial sampling ratio as $\InitRatio=10\%$ and step size as $\StepSize=3\%$, its latency returned to a sub-second level (0.96s) without any accuracy loss.

In traditional online AQP and in Biathlon, the median computed via bootstrapping may require more samples when the data distribution is discrete and uniform. \RII{Response 3.1} 
Nevertheless, this would not affect the correctness of Biathlon because, in the worst case, Biathlon would eventually sample all the data for that specific feature. However, depending on the importance of that feature and the number of features in a pipeline, Biathlon can still provide significant speedup in practice.

To study this issue, we conducted two more experiments. 
In the first experiment, 
we replaced one AVERAGE operator in the Bearing-Imbalance pipeline with a MEDIAN operator.  Different from the experiment above, 
here, we also replaced the data used to compute that MEDIAN feature with synthetic data.
The synthetic data was generated with 2,000,001 rows, containing either the value $x$ or the value $x + 100$, where $x$ is the original value from 
the AVERAGE aggregation in the unmodified Bearing-Imbalance pipeline. This approach ensures that sampling all data for MEDIAN would yield a MEDIAN value identical to the original input feature, thereby eliminating the need for model retraining. On the other hand, sampling some but not all data for MEDIAN would result in obtaining either a MEDIAN value identical to the original input feature or a MEDIAN value that is off by 100.

Since there are now only two possible values in the data for computing the MEDIAN, we treat these values as two "classes" and synthesize 2,000,001 rows of data with different "class imbalance ratios".\footnote{In machine learning, class imbalance refers to the situation when the number of instances in each class of a dataset is not equal.} The class imbalance ratio represents the number of rows with a value of $x + 100$ divided by the number of rows with a value of $x$. Thus, a class imbalance ratio close to 1 represents a pathological distribution that is discrete and uniform.

Figure \ref{fig:3.2.1} illustrates the performance and accuracy of Biathlon across different class imbalance ratios in the modified pipeline. 
The blue line shows the percentage of data sampled over the total data size.
The orange line shows the percentage of data specifically sampled for 
computing the MEDIAN among the 2,000,001 rows.
We can observe that Biathlon overall remains resilient to the class imbalance ratio until it reaches the extreme values ($\ge$ 0.95), where up to 21\% of the synthetic data for the MEDIAN feature are sampled. However, since the MEDIAN operator is not the only operator in this pipeline, the overall percentage of data sampled remains low (6.1--8.3\%), as there are other features that use way fewer samples.

Since we have replaced an input feature with the MEDIAN operator, it is no longer meaningful to measure the accuracy based on the true labels in the hold-out set.
Therefore, we measure the accuracy of Biathlon using the 
exact value predicted by the baseline as the oracle label. Notably, Biathlon's predictions on this classification pipeline 100\% matches the predictions on the exact pipeline across all class imbalance ratios.

In the second experiment, we repeated the same process with the Tick-Price regression pipeline. We chose this pipeline because it contains only one aggregation operator, hence there are no other aggregation operators to dilute the effect of the pathological data distribution.

Figure \ref{fig:3.2.2} displays the results of the experiment. We observe that Biathlon's performance continues to exhibit resilience across a wide range of class imbalance ratios. However, it indeed requires more samples earlier, at a class imbalance ratio of 0.9. Nevertheless, Biathlon can still avoid accessing the full dataset 
except when the class imbalance ratio is 1.
For this regression pipeline, we measure the absolute error between Biathlon's predictions and the predictions made by the baseline, which samples all the data. The results demonstrate that the prediction error remains close to 0 and steady across all class imbalance ratios.
}

\revision{In summary, it is important to note that the uniformity of data 
may bring some pathological cases to MEDIAN approximation.
Nonetheless, it is not specific to Biathlon but to all sampling-based AQP systems \cite{DBO, OLA, DeepOLA}.
In fact, we can observe that Biathlon consistently maintains high accuracy and offers speedup across a wide range of data distribution.
In terms of limitation, similar to traditional online AQP systems, Biathlon does not approximate top-k queries, distinct counting, and extreme statistics such as min and max.\cite{OLA,DBO,GOLA,SampleSeek,Quickr,XDB,KnowingWrong} 
For such cases,
precomputing aggregates offline, as demonstrated by RALF, can be a viable alternative. However, it is essential to remark that the feature store approach (including RALF) also has its limitations, such as the absence of error bounds or being restricted to a limited set of workloads. Therefore, we believe that the feature store caching methodology of RALF and the AQP approach of Biathlon can complement each other, opening up an intriguing avenue for future research.
}

\end{document}